\documentclass[useAMS,usenatbib]{mn2e}
\usepackage{graphicx}

\usepackage{epstopdf} 

\newcommand{\spose}[1]{\hbox to 0pt{#1\hss}}
\newcommand{\approxpropto}{\mathrel{\spose{\lower 3pt\hbox{$\sim$}}
	\raise 2.0pt\hbox{$\propto$}}}
\def\approxgt{\mathrel{\spose{\lower 3pt\hbox{$\sim$}}
	\raise 2.0pt\hbox{$>$}}}
\def\approxlt{\mathrel{\spose{\lower 3pt\hbox{$\sim$}}
	\raise 2.0pt\hbox{$<$}}}

\newcommand{\Nbody}{$N$-body}
\newcommand{\LCDM}{$\Lambda$CDM\,}
\newcommand{\Hubbleunits}{\hbox{kms}^{-1}\hbox{Mpc}^{-1}\,}

\def \hkpc{\,h^{-1} \hbox{kpc}}
\def \hMpc{\,h^{-1} \hbox{Mpc}}
\def \hMsol{\,h^{-1} \hbox{M}_{\odot}}
\def \Zsol{\,\hbox{Z}_{\odot}}

\def \h2gcm3{h^{2} \, \hbox{g}\,\hbox{cm}^{-3}}

\def \kevcm2{\, \hbox{keV} \, \hbox{cm}^{2}}
\def \cm3{\, \hbox{cm}^{-3}}

 \title[X-ray clusters in the CLEF simulation]
        {The evolution of clusters in the CLEF 
	  cosmological simulation: X-ray 
	  structural and scaling properties}
\author[S.T. Kay et al.]
            {Scott T. Kay,$^{1,2}$\thanks{E-mail: skay@astro.ox.ac.uk}
	    Antonio C. da Silva,$^{3,4}$ 
	    Nabila Aghanim,$^{4}$
	    Alain Blanchard,$^{5}$ \newauthor
	    Andrew R. Liddle,$^{2}$
	    Jean-Loup Puget,$^{4}$ 
	    Rachida Sadat$^{5}$ and
	    Peter A. Thomas$^{2}$ \\
	$^{1}$Astrophysics, Department of Physics, University of
        Oxford, Keble Road, Oxford OX1 3RH\\ 
        $^{2}$Department of Physics and Astronomy, University of
        Sussex, Falmer,  
	Brighton BN1 9QH\\
	$^{3}$Centro de Astrof\'{ii}sica da Universidade do Porto, 
	Rua das Estrelas, 4150-764 Porto, Portugal\\
	$^{4}$Institut d'Astrophysique Spatiale (IAS), B\^atiment 121,
            F-91405 Orsay; Universit\'e Paris-Sud 11 and CNRS (UMR
        8617), France\\ 
	$^{5}$Observatoire Midi-Pyr\'en\'ees, Av. Edouard Belin 14,
        31500 Toulouse, France \\ 
	}

\begin{document}

\date{This draft was generated on \today}

\pagerange{\pageref{firstpage}--\pageref{lastpage}} \pubyear{2005}

\maketitle

\label{firstpage}

\begin{abstract}
We present results from a study of the X-ray cluster population that
forms within the CLEF cosmological hydrodynamics simulation, a large
\Nbody/SPH simulation of the \LCDM cosmology with radiative cooling,
star formation and feedback.  With nearly one hundred ($kT>2$ keV)
clusters at $z=0$ and sixty at $z=1$, our sample is one of the largest
ever drawn from a single simulation and allows us to study variations
within the X-ray cluster population both at low and high redshift.
The scaled projected temperature and entropy profiles at
$z=0$ are in good agreement with recent high-quality observations of 
cool core clusters, suggesting that the simulation grossly follows the 
processes that structure the intracluster medium (ICM) in these objects. 
Cool cores are a ubiquitous phenomenon in the simulation at low and
high redshift, regardless of a cluster's dynamical state. This is at odds 
with the observations and so suggests there is still a heating mechanism 
missing from the simulation. The fraction of irregular (major
merger) systems, based on an observable measure of substructure within
X-ray surface-brightness maps, increases with redshift, but always
constitutes a minority population within the simulation.  Using a
simple, observable measure of the concentration of the ICM, which
correlates with the apparent mass deposition rate in the cluster core,
we find a large dispersion within regular clusters at low redshift,
but this diminishes at higher redshift, where strong {\it
cooling-flow} systems are absent in our simulation. Consequently, our
results predict that the normalisation and scatter of the 
luminosity--temperature relation should decrease with redshift; if
such behaviour turns out to be a correct representation of X-ray
cluster evolution, it will have significant consequences for the
number of clusters found at high redshift in X-ray flux-limited surveys.
\end{abstract}


\section{Introduction}

Clusters of galaxies, occurring at low redshifts ($z<2$), are
interesting cosmological objects as they offer a powerful yet
independent approach from other methods (such as the cosmic microwave
background, at $z \sim 1000$) for constraining cosmological
parameters. While several cosmological applications of clusters exist,
a particularly appealing method, because of its simplicity, is to
measure the variation in the cluster mass function with redshift
(e.g. \citealt{Blanchard98,Eke98b}). Since mass is known to be tightly
correlated with X-ray observables, particularly temperature
(e.g. \citealt{Finoguenov01,Arnaud05,Vikhlinin05b}), it is
straightforward in principle to convert between the two quantities.

Theoretically, such cluster scaling relations were predicted to exist,
essentially as a manifestation of the virial theorem, by
\citet{Kaiser86}. In the so-called {\it gravitational-heating}
scenario, the intracluster medium (ICM) was heated by the
gravitational collapse and subsequent virialisation of the
cluster. X-ray observations of clusters confirmed the existence of
these scaling relations (e.g. \citealt{Edge91,Fabian94} for the X-ray
luminosity--temperature relation) though they revealed two
complications.  Firstly, the slope of the observed X-ray
luminosity--temperature relation (and to a lesser extent, the
mass--temperature relation) is steeper than predicted from
gravitational heating alone, an effect shown by \citet{Ponman99} to be
due to an excess of entropy in the cores of clusters, and more so in
groups. A lot of theoretical effort has gone into understanding the
origin of the excess entropy (see \citealt{Voit05} for a recent
review). Secondly, there is an intrinsic scatter in the scaling
relations, which is particularly large for the low-redshift
luminosity--temperature relation, due to the large variations in core
luminosity \citep{Fabian94}. For cosmological studies with clusters, an
accurate statistical description of the cluster population is
warranted, as only then can robust cluster survey selection functions
be constructed. From a theoretical stand-point, cluster scaling
relations offer an additional, exciting prospect; the amount with
which these relations evolve with redshift ought to reveal information
on the nature of non-gravitational processes and cluster astrophysics
in general \citep{Muanwong06}.

The intrinsic scatter in cluster scaling relations can at least partly
be attributed to gravitational processes, as clusters themselves live
in different environments (e.g.~\citealt{Schuecker01}), although
non-gravitational processes, such as radiative cooling and heating
from galaxies, must also play a role
(e.g.~\citealt{Pearce00,McCarthy04,Balogh06}).  Cosmological
simulations of the cluster population are the most accurate method by
which to characterise the statistical properties of clusters, as they
include an accurate treatment of the non-linear gravitational dynamics
and merging processes, as well as allowing non-gravitational physics
to be incorporated self-consistently. Early attempts focused on the
simplest model for the gas, a non-radiative ICM, which was
successfully shown to reproduce the simple, self-similar, scalings
expected from the gravitational-heating model
(e.g.~\citealt{Navarro95,Evrard96,Bryan98,Eke98a,Muanwong02}).

Additional non-gravitational processes have also been studied within
simulations, and various mechanisms have been proposed to explain the
similarity breaking, such as {\it preheating}
(e.g.~\citealt{Navarro95,Bialek01,Borgani02}), radiative cooling
(e.g.~\citealt{Pearce00,Muanwong01,Muanwong02,Dave02,Motl04,Kravtsov05})
or both (e.g. \citealt{Muanwong02}). Recently, attention has shifted
to more realistic models which attempt to directly couple feedback
(local heating from galaxies) with cooling and star formation
(e.g.~\citealt{Valdarnini03,Tornatore03,Kay2003,Borgani04,
Kay2004a,Kay2004b,Ettori04}).

Together with progress in the development of these non-gravitational
models, the advance in both simulation codes and computer hardware is
now allowing larger simulations with reasonable resolution to be
performed. We are now beginning to resolve sufficient numbers of
clusters to start making quantitative predictions at all appropriate
redshifts for the cluster population.  The CLEF-SSH (CLuster Evolution
and Formation in Supercomputer Simulations with Hydrodynamics)
collaboration has been set up to take advantage of this new era in
numerical modelling, by performing large simulations of the cluster
population. Our first simulation, known as the CLEF simulation, is a
large ($N=2\times 428^3$) $N$-body/SPH simulation of the $\Lambda$CDM
cosmology, within a $200\hMpc$ box, and includes a model for radiative
cooling and energy feedback from galaxies. This simulation is a
similar size to the one performed by \citet{Borgani04}, but uses a
different feedback model.  In \citet{Kay2005}, hereafter Paper~I, we
presented a small selection of results at $z=0$ from the CLEF
simulation. For this paper, we have performed a more detailed analysis
of the same cluster population, and present results for a range of
redshifts from $z=0$ to 1, focusing on the effects of dynamical
activity and the strength of cooling cores. A companion paper (da
Silva et al., in preparation) presents results for the
Sunyaev--Zel'dovich effect properties of the CLEF cluster population.

The rest of this paper is organised as follows.  In
Section~\ref{sec:sim} we summarise details of the CLEF simulation and
detail our method for creating cluster catalogues, maps and profiles.
The internal structure of clusters, and how it depends on dynamical
regularity and the properties of the core, is the focus of
Section~\ref{sec:structure}.  Section~\ref{sec:scale} then draws on
these results to investigate the evolution of key cluster scaling
relations with redshift. We discuss our results in
Section~\ref{sec:disc} and summarise our conclusions in
Section~\ref{sec:conc}.

\section{The CLEF simulation}
\label{sec:sim}

The CLEF simulation is a large ($N=2\times 428^3$ particles within a
$200 \hMpc$ comoving box) cosmological simulation of structure
formation, incorporating both dark matter and gas. Below we describe
the procedure used to generate the simulation data and how the
clusters were identified within these data to create X-ray
temperature-limited samples from redshifts $z=0$ to $z=1$.

\subsection{Simulation details}

For the cosmological model, we adopted the spatially-flat \LCDM
cosmology, setting the following values for cosmological parameters:
matter density parameter, $\Omega_{\rm m}=0.3$; cosmological constant,
$\Omega_{\Lambda}=\Lambda/3H_0^2=0.7$; baryon density parameter,
$\Omega_{\rm b}=0.0486$; Hubble constant, $h=H_0/100\Hubbleunits=0.7$;
primordial power spectrum index, $n=1$ and power spectrum
normalisation, $\sigma_8=0.9$.  These values were chosen to be
consistent with the results from {\it WMAP} first year data
\citep{Spergel03}.

Initial conditions were generated for a cube of comoving length $200
\hMpc$ at redshift, $z=49$. The cube was populated with two
interleaving grids of $428^3$ particles, one grid representing the
dark matter and one representing the gas; the particle masses were
thus set to $m_{\rm dark}=7.1\times 10^{9}\hMsol$ and $m_{\rm
gas}=1.4\times 10^{9}\hMsol$ for the dark matter and gas respectively.
Initial particle displacements and velocities were then computed from
a transfer function generated using the CMBFAST code
\citep{Seljak96}. The initial temperature of the gas was set to
$T=100$\,K, significantly lower than the range of temperatures typical
of overdense structures resolved by the simulation.

The initial conditions were evolved to $z=0$ using a version of
the GADGET2 \Nbody/SPH code \citep{Springel05}, modified to include
additional physical processes (radiative cooling, star formation and
energy feedback; see below).  Gravitational forces were calculated
using the Particle-Mesh algorithm on large scales (using a $512^3$
FFT) and the hierarchical tree method on small scales.  The
(equivalent) Plummer softening length was set to $\epsilon=20\hkpc$,
fixed in comoving co-ordinates, thus softening the Newtonian force law
below a comoving separation, $x_{\rm min}=2.8\,\epsilon=56\hkpc$.

Gas particles were additionally subjected to adiabatic forces, and an
artificial viscosity where the flow was convergent, using the
entropy-conserving version of SPH \citep{Springel02}, the default
method in GADGET2. Additionally, we allowed gas particles with
$T>10^{4}$K to cool radiatively, using the isochoric cooling
approximation suggested by \citet{Thomas92}. Tabulated cooling rates
were taken from \citet{Sutherland93}, assuming an optically thin
$Z=0.3\Zsol$ plasma in collisional ionisation equilibrium (a good
approximation to the intracluster medium out to at least $z=1$;
\citealt{Tozzi03}).

\subsection{Feedback}

We have also attempted to follow crudely the large-scale effects of
galactic outflows (feedback) in the simulation, to regulate the
cooling rate and to inject non-gravitational energy into the gas. We
adopted the {\it Strong Feedback} model of \citet{Kay2004a}, hereafter
K2004a, as it was shown there to approximately reproduce the observed
excess entropy in groups/clusters, both at small and large radii (see
also Paper~I).  We only give a brief summary of the model details
here.

First of all, cooled gas is identified with overdensity $\delta>100$,
hydrogen density $n_{\rm H}>10^{-3}\cm3$, and temperature $T<12,000$K.
For each cooled gas particle, a random number, $r$, is drawn from the
unit interval and the gas is reheated if $r<f_{\rm heat}$, where
$f_{\rm heat}=0.1$ is the reheated mass fraction parameter. Reheated
gas is given a fixed amount of entropy,\footnote{We define entropy as
$S=kT\,(\rho/\mu m_{\rm H})^{1-\gamma}$, where $\gamma=5/3$ is the
ratio of specific heats for a monatomic ideal gas and $\mu m_{\rm
H}=0.6$ is the mean atomic weight of a fully-ionised plasma.} $S_{\rm
heat}=1000\kevcm2$, corresponding to a minimum thermal energy of $\sim
17\,$keV at the star-formation density threshold. Such a high thermal
energy (compared with typical cluster virial temperatures) means that
the reheated gas is supersonic and is thus distributed through viscous
interactions and shocks in the ICM.  This not only regulates the
star-formation rate in the cluster \citep{Balogh01}, it also prevents
significant build-up of low-entropy material in the cluster core
\citep{Kay2003,Kay2004a}. Our model could thus be perceived as a crude
representation of local accretion-triggered heating by stars and
active galactic nuclei (although feedback from the latter does not
necessarily have to follow the star-formation rate, as is done here).

Gas particles that are not reheated are instead converted to
collisionless star particles.  Although the model does not treat star
formation (which occurs in regions with much higher gas densities,
$n_{\rm H} \approxgt 0.1$) accurately, this premature removal of
low-pressure material from the gas phase saves computational effort as
these particles generally have the shortest timesteps. Furthermore, it
helps to alleviate the difficulty that standard SPH has in resolving
the sharp interface between hot and cold phases \citep{Pearce00}.

\subsection{Cluster identification}

The CLEF simulation produced a total of 72 snapshots of the particle
data, at time intervals optimised for producing mock lightcones (da
Silva et al., in preparation). Only the 25 lowest-redshift snapshots
are used in this paper, ranging from $z=0$ to 1; at higher redshift the
number of clusters becomes prohibitively small. We used these
snapshots to produce cluster catalogues (mass, radius and various
other properties), maps and profiles.

Catalogues were generated using a similar procedure to that adopted by
\citet{Muanwong02}. Briefly, groups of dark matter particles were
identified using the friends-of-friends (FoF) algorithm
\citep{Davis85}, setting the dimensionless linking-length to
$b=0.1$. Spheres were then grown around the particle in each group
with the most negative gravitational potential, until the enclosed
mass equalled a critical value
\begin{equation}
M_{\Delta}(<R_{\Delta}) = {4 \over 3} \pi R_{\Delta}^3 \, \Delta \,
\rho_{\rm cr}(z), 
\label{eqn:mdelta}
\end{equation}
where $\Delta$ is the density contrast, $\rho_{\rm cr}(z)=(3H_0^2/8\pi
G)E(z)^2$ is the critical density and $E(z)^2=\Omega_{\rm
m}(1+z)^3+1-\Omega_{\rm m}$ for a flat universe \citep{Bryan98}.
Cluster catalogues for a variety of density contrasts were
constructed, including the virial value ($\Delta \sim 100$ at $z=0$),
taken from equation~(6) in \citet{Bryan98}. The virial radius was used
to find overlapping pairs, and the least massive cluster in each pair
was discarded from the catalogues.  For nearly all of the results
presented in this paper, we use a catalogue with $\Delta=500$, as this
is the smallest density contrast typically accessible to current X-ray
observations.

We initially selected all clusters with at least 3000 particles within
$R_{500}$, corresponding to a lower mass limit of $M_{500}=2.5\times
10^{13}\hMsol$. This limit is low enough that our temperature-selected
sample (below) is comfortably a complete subset of this sample at all
redshifts studied.  At $z=0$ we have 641 clusters in our mass-limited
sample, decreasing to 191 clusters at $z=1$. This is comparable to the
numbers found in the simulation performed by \citet{Borgani04}, who
also used GADGET2 but with a different prescription for cooling, star
formation and feedback than used here.

\subsection{Spectroscopic-like temperature}

Observational samples of X-ray clusters are usually limited in flux or
temperature, where the latter is measured by fitting an isothermal
plasma model to the observed spectrum of the cluster X-ray
emission. Theoretical models of X-ray clusters commonly use an
emission-weighted temperature to estimate the spectral temperature of
a cluster.  For particle-based simulations, this is done using the
formula
\begin{equation}
T = {\sum_i \, w_i T_i \over \sum_i \, w_i},
\label{eqn:ewt}
\end{equation}
where $w_i=m_i n_i \Lambda(T_i,Z)$ is the weight given to each hot
($T_i>10^5$ K) gas particle $i$, $m_i$ its mass, $n_i$ its density,
$T_i$ its temperature and $\Lambda(T_i,Z)$ the cooling function,
usually for emission within the X-ray energy band
(e.g. \citealt{Muanwong02,Borgani04}). In the bremsstrahlung regime
$\Lambda \propto T^{1/2}$ and so the hottest, densest particles are
given the most weight.

\citet{Mazzotta04} applied the same method used by observers to
measure the spectroscopic temperature of simulated clusters, and found
that it was always lower than $T_{\rm ew}$. For the bremsstrahlung
regime ($kT>2$ keV), they suggested a more accurate measure, known as
the {\it spectroscopic-like} temperature, with weight $w_i=m_i n_i
T_i^{-3/4}$. This estimator gives the coldest, densest particles more
weight. We adopt this estimator in this paper (summing over gas
particles within $R_{500}$ with $kT_i>0.5$ keV, the typical lower
energy limit of an X-ray band) to create a temperature-limited
($kT_{\rm sl}>2$ keV) sample of clusters at all redshifts.  This
reduces the number of clusters to 95 at $z=0$, decreasing to 57 at
$z=1$. While this is one of the largest temperature-selected cluster
sample drawn from a single simulation, we note that the dynamic range
is still quite small. Nearly all clusters have $T_{\rm sl} \sim 2-4$
keV at all redshifts (the median temperature stays approximately
constant with redshift at $T_{\rm sl} \sim 2.5$ keV) and our hottest
cluster at $z=0$ has $T_{\rm sl}=7.3$ keV.

\begin{figure}
\centering
\includegraphics[width=8.5cm]{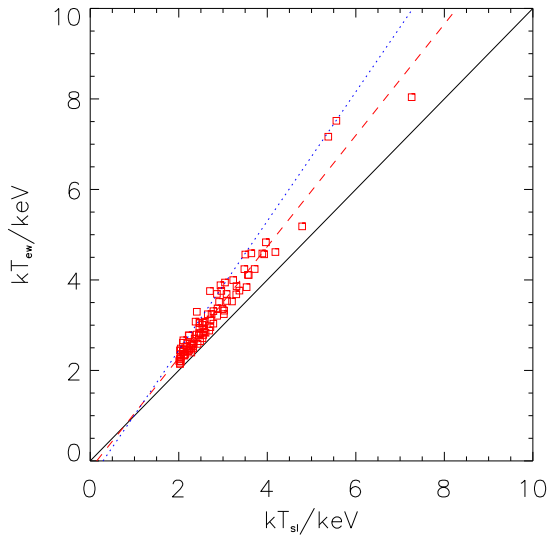} 
\caption{X-ray emission-weighted temperature, $T_{\rm ew}$, plotted
against spectroscopic-like temperature, $T_{\rm sl}$, for clusters
with $kT_{\rm sl}>2$ keV. The solid line corresponds to $T_{\rm
ew}=T_{\rm sl}$ and the dashed line is the best-fit straight line to
our data.  The dotted line is the best fit to the clusters studied by
Rasia et al. (2005).}
\label{fig:tewtsl}
\end{figure}

Fig.~\ref{fig:tewtsl} compares $T_{\rm sl}$ to $T_{\rm ew}$ (for a
$0.5-10$ keV band) for our temperature-limited sample at $z=0$. As was
found by \citet{Rasia05}, whose sample mainly consisted of the
\citet{Borgani04} clusters, $T_{\rm sl}$ and $T_{\rm ew}$ differ
by as much as 20 per cent.  Rasia et
al. found $kT_{\rm sl}=0.7 kT_{\rm ew} + 0.3$, whereas we find $kT_{\rm
sl}=0.8 kT_{\rm ew} + 0.1$, similar to, although slightly steeper than,
their result (see also \citealt{Kawahara07}).

\subsection{Cluster maps and projected profiles}

\begin{figure*}
\centering
\includegraphics[width=17cm]{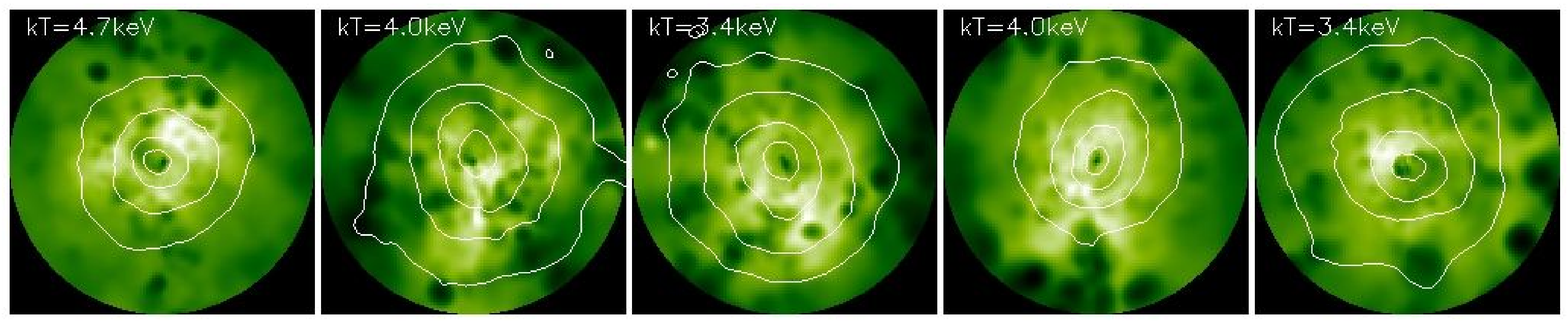}
\includegraphics[width=17cm]{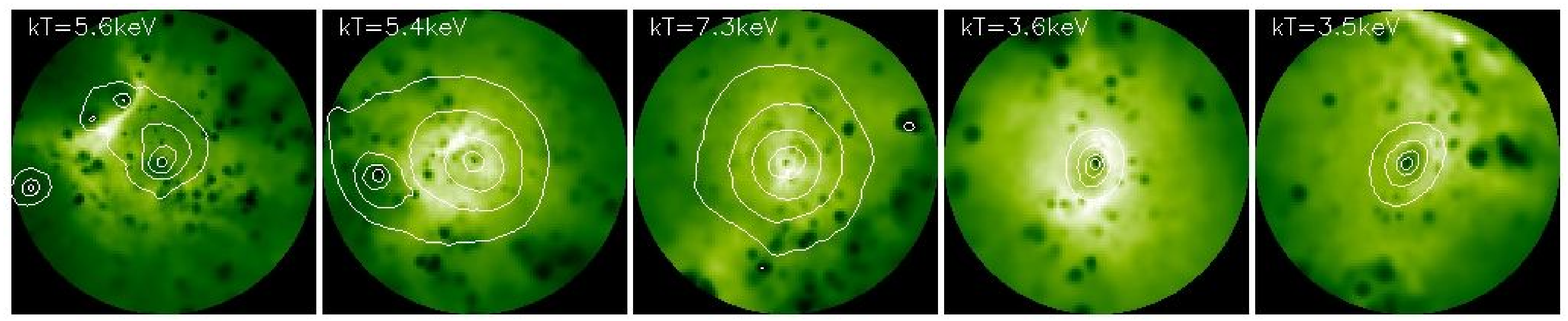}
\caption{Spectroscopic-like temperature maps of the 5 most massive
clusters (in order of decreasing mass, left to right) at $z=1$ (top
panels) and $z=0$ (bottom panels). The spectroscopic-like temperature
is given in each panel.  Surface-brightness contours (normalised to
the maximum value) are overlaid; adjacent contours correspond to a
difference in surface brightness of a factor of 4. Images are centred
on the maximum surface-brightness pixel and are shown out to $R_{500}$.}
\label{fig:sltimages}
\end{figure*}

Cluster maps were produced using a similar procedure to that discussed
in \citet{Onuora03}; in essence, values at each pixel are the sum of
smoothed contributions from particles, using the same spline kernel as
used by the GADGET2 code. Centred on each cluster, only particles
within a cube of half-length, $l=4R_{500}$, were considered, i.e.  out
to approximately twice the virial radius in each orthogonal
direction. For this paper, we computed bolometric surface brightness
(although the emission is predominantly thermal bremsstrahlung in the
X-ray) and spectroscopic-like temperature maps. Projected temperature
and azimuthally-averaged surface-brightness profiles were also
computed, averaging particles within cylindrical shells, centred on
the pixel with the highest surface brightness.

Fig.~\ref{fig:sltimages} illustrates spectroscopic-like temperature
maps of the 5 most massive clusters each at $z=0$ and $z=1$, out to
a radius, $R_{500}$.  As was found by \citet{Onuora03} and
\citet{Motl04}, there is a large amount of temperature structure
within each cluster, particularly cold spots due to cool, low-entropy
gas trapped within infalling sub-clusters. The intensity scale is
defined by the minimum and maximum temperature; the dynamic range is
typically an order of magnitude ($T_{\rm max}/T_{\rm min}$), with
maximum temperatures being around twice that of the mean.

\section{Cluster structure}
\label{sec:structure}

In this section we present the structural properties of the CLEF
clusters, comparing to observational data where appropriate.

\subsection{X-ray temperature bias}
\label{subsec:clustemp}

As discussed in the previous section, the X-ray temperature of the ICM
is biased to regions of high density. Cooling and heating processes
are generally most efficient there, so the X-ray temperature of a
cluster is not necessarily an accurate measure of the depth of the
underlying gravitational potential well, even if the system is
virialised and approximately in hydrostatic equilibrium.

We investigate any such temperature bias in our simulation by
comparing $T_{\rm sl}$ to the dynamical temperature, $T_{\rm dyn}$, 
through the standard quantity, 
$\beta_{\rm spec}=T_{\rm dyn}/T_{\rm sl}$. The dynamical temperature,
\begin{equation}
kT_{\rm dyn} = 
{
\sum_{i,{\rm gas}} \, m_i k T_i + \alpha \sum_i \, \frac{1}{2}m_i v_i^2
\over 
\sum_i \, m_i
},
\label{eqn:tdyn}
\end{equation}
where $\alpha=(2/3)\mu m_{\rm H} \sim 6.7\times 10^{-25}\,{\rm g}$,
assuming the ratio of specific heats for a monatomic ideal gas,
$\gamma=5/3$, and the mean atomic weight of a zero metallicity gas,
$\mu m_{\rm H}=10^{-24}$ g. The first sum in the numerator runs over
gas particles and the second sum over all particles, of mass
$m_i$, temperature $T_i$ and speed $v_i$ in the centre of momentum
frame of the cluster.

\begin{figure}
\centering
\includegraphics[width=8cm]{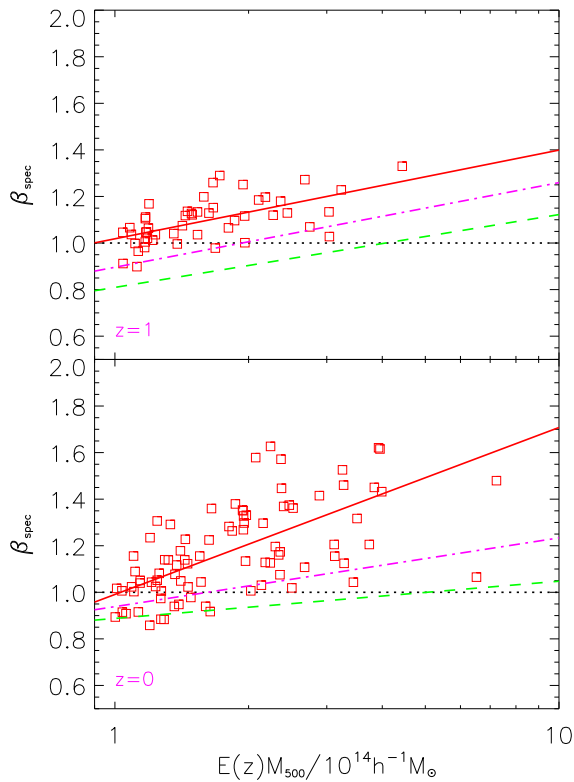} 
\caption{Ratio of dynamical temperature to spectroscopic-like
temperature, $\beta_{\rm spec}=T_{\rm dyn}/T_{\rm sl}$, 
versus scaled mass, $E(z)M_{500}$, at $z=1$ and
$z=0$. The solid line is the best-fit relation to the data,
while the dot-dashed (dashed) lines are best-fit relations 
when $T_{\rm sl}$ is replaced by the hot gas mass-weighted temperature 
excluding (including) bulk kinetic motions.}
\label{fig:trat}
\end{figure}

Fig.~\ref{fig:trat} illustrates $\beta_{\rm spec}$ for each cluster at
$z=1$ and $z=0$, versus its scaled mass, $E(z)M_{500}$. Only clusters
with $E(z)M_{500}>10^{14}\hMsol$ are selected, producing similar
numbers to our temperature-selected samples at both redshifts.  There
is a clear positive correlation between $\beta_{\rm spec}$ and
$E(z)M_{500}$, with $\beta_{\rm spec}>1$ for most clusters (i.e.
$T_{\rm sl}<T_{\rm dyn}$) at low and high redshift.  The
spectroscopic-like temperature is a biased tracer of the gravitational
potential for three reasons. Firstly, cool dense gas is weighted more
than less dense material, as discussed in the previous section. This
effect can be seen in the figure by comparing $\beta_{\rm spec}$ to
the best-fit relation when $T_{\rm sl}$ is replaced by the hot gas
mass-weighted temperature (dot-dashed line).  Secondly, some of the
energy of the gas is in macroscopic kinetic energy, as can be deduced
from comparing the dot-dashed to the dashed line, where in the latter
case, $T_{\rm sl}$ is replaced by the temperature when
equation~(\ref{eqn:tdyn}) is applied to only the hot gas. Finally,
feedback heats the gas, particularly in low mass clusters (the dashed
line shows that $\beta_{\rm spec}<1$ for most clusters, i.e. 
the gas has more specific energy than the dark matter).

\subsection{Baryon fractions}

\begin{figure}
\centering
\includegraphics[width=8cm]{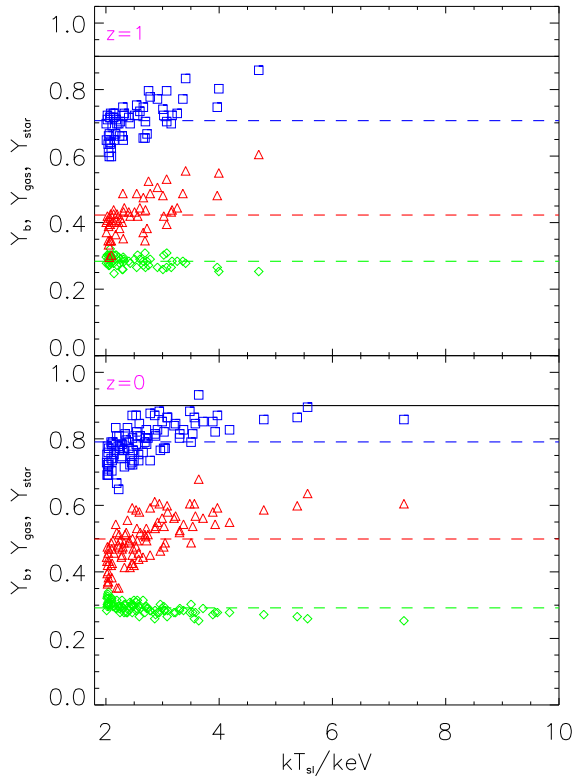}
\caption{Baryon (squares), gas (triangles) and star (diamonds)
fractions versus spectroscopic-like temperature at $z=1$ and
\mbox{$z=0$}.  Horizontal dashed lines illustrate mean values.  The
solid line is the average value measured by \citet{Kay2004b} for their
{\it non-radiative} clusters.}
\label{fig:bfrac}
\end{figure}

We also examine the segregation of baryonic mass into gaseous (ICM)
and galactic (collisionless) components within each cluster. Gas and
baryon fraction profiles for this model have already been studied by
\citet{Kay2004b}, hereafter K2004b, who showed that the baryon
fraction profiles were in good agreement with observations but that
too much of the gas had turned into stars (the normalisation of the
gas fraction profile is as low as 50 per cent of the observed
profile). Here we examine the behaviour of the baryon/gas/star
fractions with temperature and redshift.

Fig.~\ref{fig:bfrac} illustrates baryon fractions normalised to the
global value, $Y_{\rm b}=f_{\rm b}/(\Omega_{\rm b}/\Omega_{\rm m})$
within $R_{500}$, for each cluster at $z=1$ and $z=0$. Most of our
clusters have $kT_{\rm sl}<5$~keV, where there is a strong trend in
increasing baryon fraction with temperature, as feedback can heat 
and expel more gas in smaller clusters. At high temperature, a few systems 
at $z=0$ are consistent with the mean
value ($\sim 0.9$) found by K2004b for their non-radiative
clusters. Overall, the mean baryon fraction increases by 8 per cent
between the two redshifts, from 0.71 at $z=1$ to 0.79 at
$z=0$. Similarly, the mean hot gas fraction increases from 0.42 to
0.49 over the same redshift range. \citet{Ettori04} also found the gas
fractions to weakly decrease with redshift in their simulated
clusters, albeit with higher values than found here.

The star fraction is a very weak function of both temperature and
redshift, with a mean value of 0.28 at $z=0$ and 0.29 at $z=1$. Just
under 40 per cent of the baryons within $R_{500}$ have condensed and
formed stars in our simulation, at all redshifts; a value that only
decreases to about 30 per cent at the virial radius. Observations
indicate a value of about 10 to 15 per cent, significantly lower than
in our clusters \citep{Lin03}. Thus, as found by K2004b, our feedback
model has not been effective enough at limiting the overcooling of
baryons in clusters, as was also found by \citet{Ettori06}. However,
the global star fraction is only 13 per cent at $z=0$ (and 9 per cent
at $z=1$), just slightly larger than the observed value of 5 to 10 per
cent (e.g.~\citealt{Balogh01}).

\subsection{Regularity}
\label{subsec:sub}

\begin{figure}
\centering
\includegraphics[width=8cm]{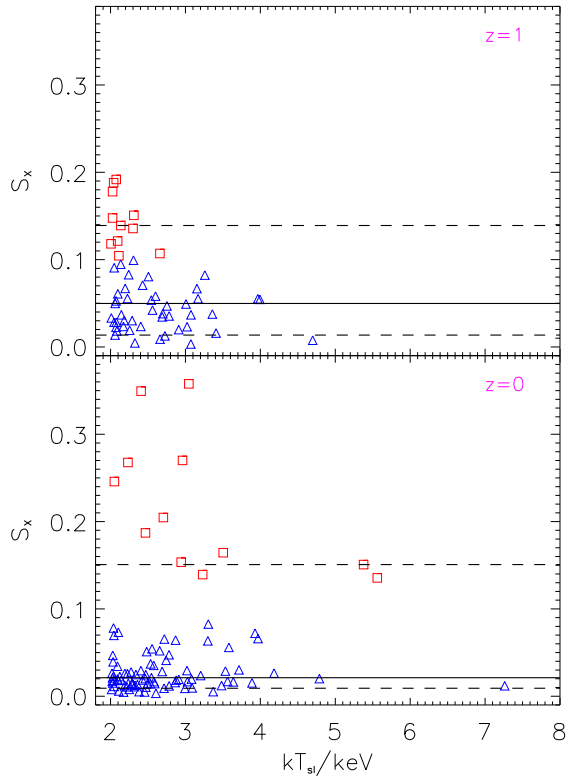} 
\caption{Substructure statistic, $S_{\rm X}$, versus
spectroscopic-like temperature for clusters at
$z=1$ and $z=0$. Triangles illustrate
regular clusters with $S_{\rm X}\le 0.1$ and squares irregular
clusters with $S_{\rm X}>0.1$. The solid horizontal line is the median
$S_{\rm X}$ and the dashed lines the 10 and 90 percentiles.}
\label{fig:subtsl}
\end{figure}

\begin{figure*}
\centering
\includegraphics[width=17cm]{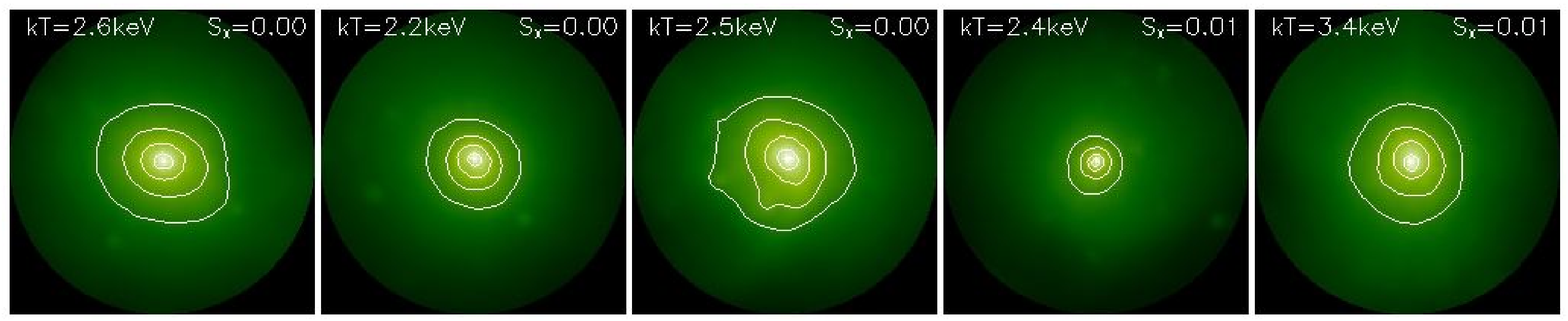}
\includegraphics[width=17cm]{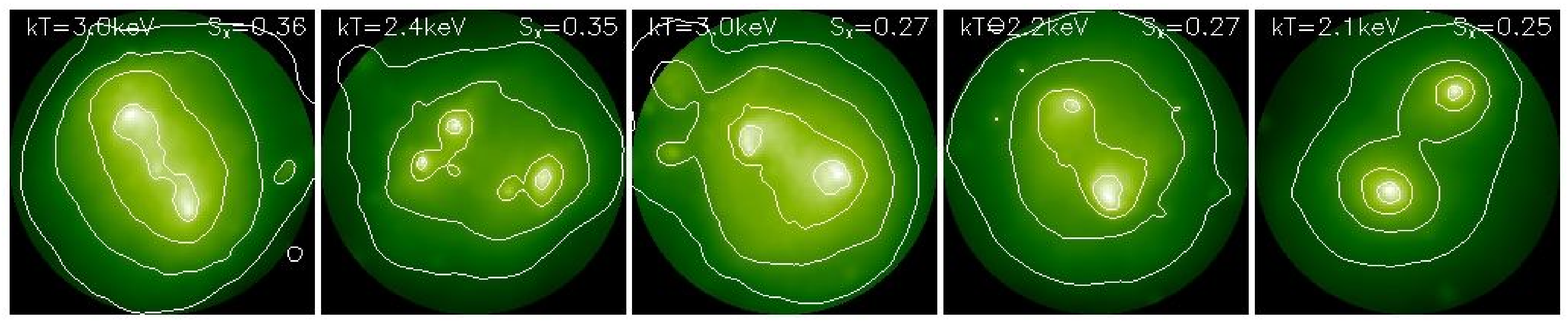}
\caption{Bolometric surface-brightness maps of the 5 clusters with the
lowest (top panels) and 5 with the highest (bottom panels) substructure
statistic ($S_{\rm X}$) at $z=0$. In this case, clusters are centred on 
the surface-brightness centroid.}
\label{fig:subimages}
\end{figure*}

Hierarchical models of structure formation predict that substructures
in clusters should be commonplace, as clusters are the latest result
of a series of mergers of smaller systems, and have dynamical times
($\sim 1$ Gyr) that are a significant fraction of the age of the
Universe.  Indeed, substructure is frequently observed in clusters and
dynamical activity has been quantified using various techniques (e.g.
\citealt{Jones92,Mohr93,Buote95,Buote96,Crone96,Schuecker01,Jeltema05}).

We use a simple measure of substructure in our cluster
surface-brightness maps, using the centroid-shift method similar to
that suggested by \citet{Mohr93}
\begin{equation}
S_{\rm X}={|{\bf R}_{\Sigma,{\rm max}}-{\bf R}_{\Sigma, \rm cen}|
  \over R_{500}}, 
\label{eqn:sub2d}
\end{equation}
where ${\bf R}_{\Sigma,{\rm max}}$ is the position of the pixel with
maximum surface brightness (taken to be the centre of the cluster) and
${\bf R}_{\Sigma, \rm cen}$ is the surface-brightness centroid.
\citet{Thomas98} used a similar method, based on the 3D total mass
distribution, and showed that this did as well, or better, than other
more sophisticated measures of substructure in simulated clusters.

Fig.~\ref{fig:subtsl} illustrates $S_{\rm X}$ versus $T_{\rm sl}$ for
our temperature-limited sample of clusters at $z=1$ (top panel) and
$z=0$ (bottom panel). It is evident that the range of $S_{\rm X}$
values at fixed temperature is large: the 10 and 90 percentiles of
each distribution, shown as horizontal dashed lines, vary from $\sim
0.01$ to $\sim 0.15$. Inspection of surface-brightness maps (see
Fig.~\ref{fig:subimages}) reveals that clusters with the largest
$S_{\rm X}$ appear dynamically disturbed and are therefore undergoing
a major merger. We choose to divide our sample into {\it irregular}
clusters with $S_{\rm X}>0.1$ and {\it regular} clusters otherwise.
We note that this division is somewhat arbitrary and only serves to
provide us with a means to compare the most disturbed clusters at each
redshift to the rest of the sample. The longer tail in the $S_{\rm X}$
distribution to high values at $z=0$ exacerbates the difference between 
the two sub-populations relative to those at $z=1$ (the length of the
tail itself changes from redshift to redshift). 

There is no significant trend in $S_{\rm X}$ with temperature, within
the limited dynamic range of our sample. However, there is a 
trend in $S_{\rm X}$ with redshift: the median value at $z=1$ is
almost a factor of 2 higher than at $z=0$. In other words, clusters tend 
to be less regular at higher redshift. 

The increase in dynamical activity with redshift in our simulated
cluster population is qualitatively consistent with the recent result
of \citet{Jeltema05}, who used the more complex power ratios
\citep{Buote95} to measure dynamical activity in a sample of low- and
high-redshift clusters observed with {\it Chandra}.

\subsection{Temperature and surface-brightness profiles}

Surface-brightness and projected temperature profiles are now
regularly observed for low-redshift clusters with {\it XMM-Newton} and
{\it Chandra} (e.g. \citealt{Arnaud05,Vikhlinin05a,Piffaretti05,Zhang06,
Pratt07}).  
These are key
observable quantities, as 3D density and temperature information can
be extracted from these measurements through deconvolution techniques.
This allows the thermodynamics of the ICM to be studied, as well as
the total mass distribution to be calculated (assuming hydrostatic
equilibrium).

\begin{figure}
\centering
\includegraphics[width=8cm]{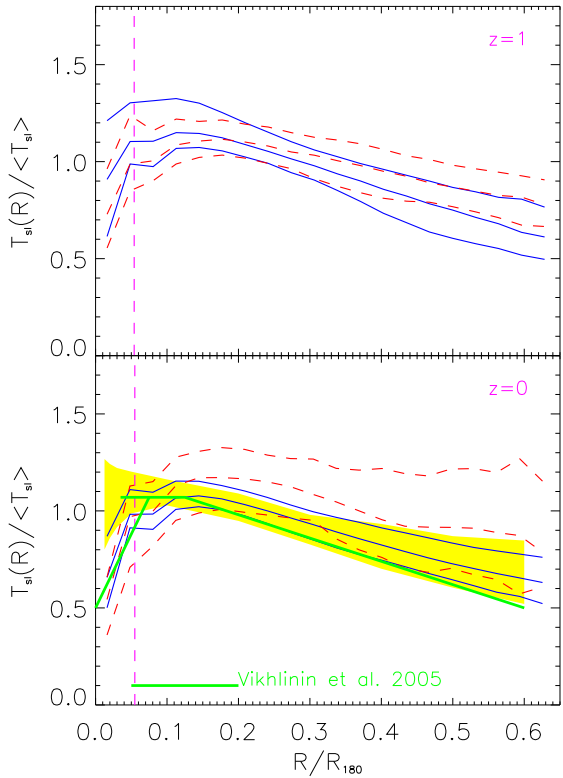}
\caption{Scaled projected spectroscopic-like temperature profiles at
$z=1$ and $z=0$. Solid curves are median
and 10/90 percentiles for regular clusters, and dashed curves
for irregular clusters. The vertical dashed line illustrates the
median scaled softening radius (i.e. where the gravitational force
becomes softer than Newtonian).  The thick solid lines (with zero and
negative gradient) are fits to the average observed temperature
profile of cool core clusters, as measured by \citet{Vikhlinin05a}; 
the inner line is a rough fit to their data to illustrate the cool core.
The shaded region encloses the mean and $1\sigma$ standard deviation 
temperature profile for a representative sample of nearby clusters by 
\citet{Pratt07}.}
\label{fig:projtprof}
\end{figure}

In Fig.~\ref{fig:projtprof} we present scaled projected
spectroscopic-like temperature profiles for our regular and irregular
clusters at $z=1$ and $z=0$. As stated previously, each cluster
(including irregular objects, where all emission is included in our
analysis) is centred on the pixel with maximum
surface brightness.  For consistency with the observational data, each
temperature profile is normalised to the average spectroscopic-like
temperature, $\left< T_{\rm sl} \right>$, between projected radii
of $50 \hkpc$ and $R_{500}$. Projected radii are then re-scaled
to $R_{180}$ using the formula $R_{180}=1.95\sqrt{k\left< T_{\rm sl}
\right>/10{\rm keV}}/E(z)\hMpc$, originally derived from numerical
simulations by \citet{Evrard96}. 

As was found by K2004b, the median profile rises sharply from the
centre outwards, peaks at $\sim 0.1 R_{180}$, then gradually declines
at larger radii. The inner rise, where the density is largest, is due
to radiative cooling of the gas, while the outer decline is a generic
prediction of the $\Lambda$CDM model (e.g. \citealt{Eke98a}).  It is
interesting to note that the shape of the profile for irregular
clusters is flatter than for the regular majority, beyond the peak.
This is due to the second, infalling, object, which compresses and
heats the gas. We also note that the temperature profiles at $z=1$ are
very similar to those at $z=0$, and so a cool core is established in
the cluster early on.

\citet{Vikhlinin05a} recently determined the projected temperature
profile for a sample of 11 low-redshift cool core clusters observed
with {\it Chandra}. The shape of their profile is very similar, albeit
slightly steeper at large radii, to that of our regular clusters; a
rough fit, as supplied by the authors, is shown in
Fig.~\ref{fig:projtprof} as thick solid lines. \citet{Pratt07} performed
a similar study with {\it XMM-Newton}, for a sample of 15 clusters
(including non-cool core systems); their result is shown in the figure
as the shaded region. Interestingly, \cite{Pratt07} find a similar decline
at large radius to our regular clusters but the temperature does not 
drop as sharply in the centre (even for those clusters with coolest cores).

The presence of cool cores at both low and high redshift in our simulation 
is in qualitative agreement with the findings of \citet{Bauer05}, who
measured central cooling times for a sample of $z=0.15-0.4$ clusters
observed with {\it Chandra} and found their distribution to be very
similar to that for a local sample.

\begin{figure}
\centering
\includegraphics[width=8cm]{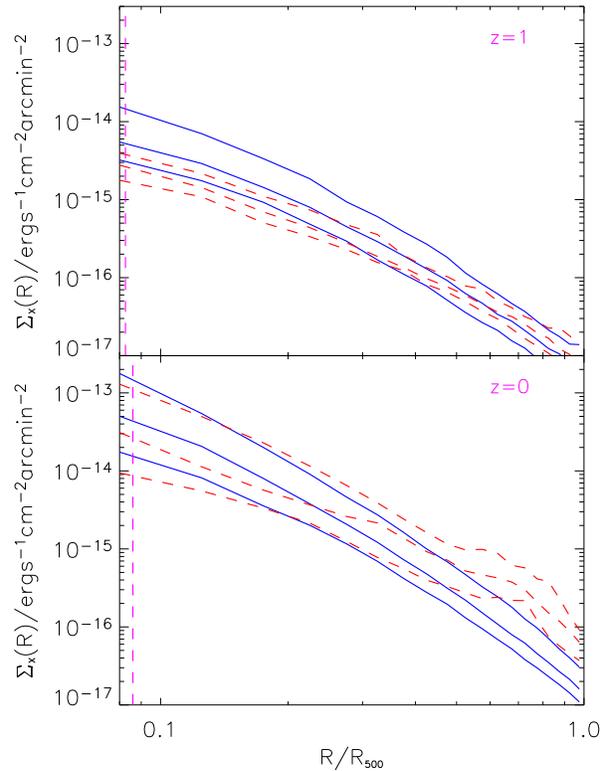}
\caption{Bolometric surface-brightness profiles at $z=1$ and $z=0$.
Again, solid curves are median and 10/90 percentile values for regular
clusters, and dashed curves for irregular clusters. The vertical
dashed line marks the median force resolution, $\left<
2.8\epsilon/R_{500}\right>$.}
\label{fig:sbprof}
\end{figure}

Bolometric surface-brightness profiles are presented in
Fig.~\ref{fig:sbprof}.  At both redshifts, it is clear that there is a
larger dispersion between clusters in the core than at the outskirts,
particularly at $z=0$. The irregular clusters have flatter profiles
than the regular clusters and a bump can be seen at large radius, due to
the core of the second object.

\begin{figure}
\includegraphics[width=9cm]{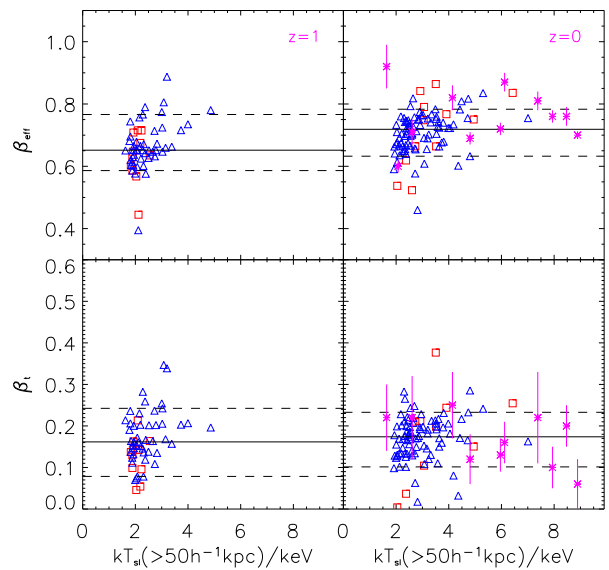}
\caption{Effective slopes of gas density ($\beta_{\rm eff}$) and
temperature ($\beta_{\rm t}$) profiles at $R_{500}$, versus
core-excised spectroscopic-like temperature, for clusters at $z=1$ and
$z=0$.  Triangles are regular clusters and squares irregular clusters.
Data points with error bars are from \citet{Vikhlinin05b}.}
\label{fig:sloper500tsl}
\end{figure}

We also calculate density and temperature gradients for our clusters,
as is needed for cluster mass estimates (Section~\ref{subsec:mest}).
Following \citet{Vikhlinin05b}, we define $\beta_{\rm eff}=-(1/3)d \,
\ln \rho / d \ln r$ and $\beta_{\rm t}=-(1/3)\, d \ln T / d \ln r$ to
represent 3D density and temperature gradients
respectively. Fig.~\ref{fig:sloper500tsl} shows these values for our
clusters at $R_{500}$, plotted against temperature. Results at $z=0$
are overplotted with the {\it Chandra} data from
\citet{Vikhlinin05b}. In general, the agreement between our results
and the observations is very good; median values are $\beta_{\rm
eff}=0.76$ and $\beta_{\rm t}=0.19$ respectively. At $z=1$ the median
values change very little (0.73 and 0.18).

\subsection{Core structure parameters}
\label{subsec:structparam}

\begin{figure*}
\centering
\includegraphics[width=17cm]{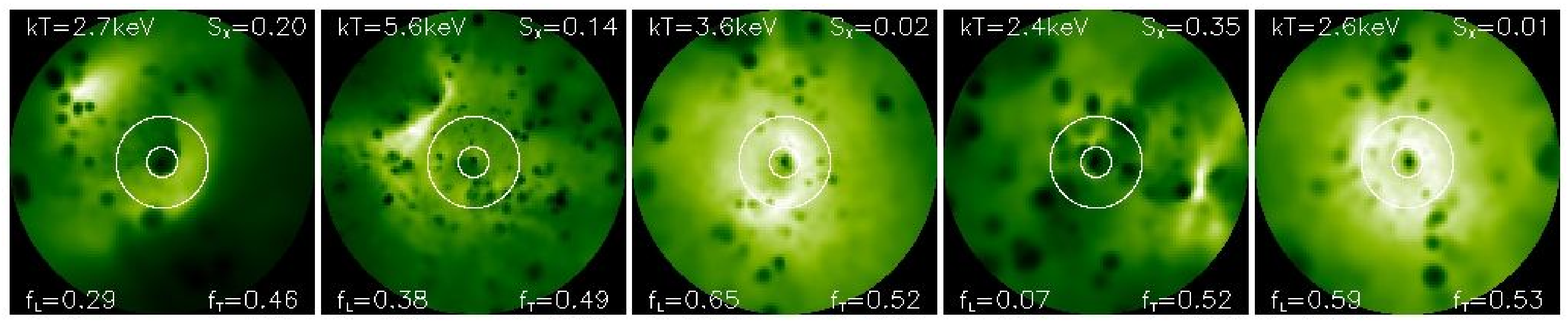}
\includegraphics[width=17cm]{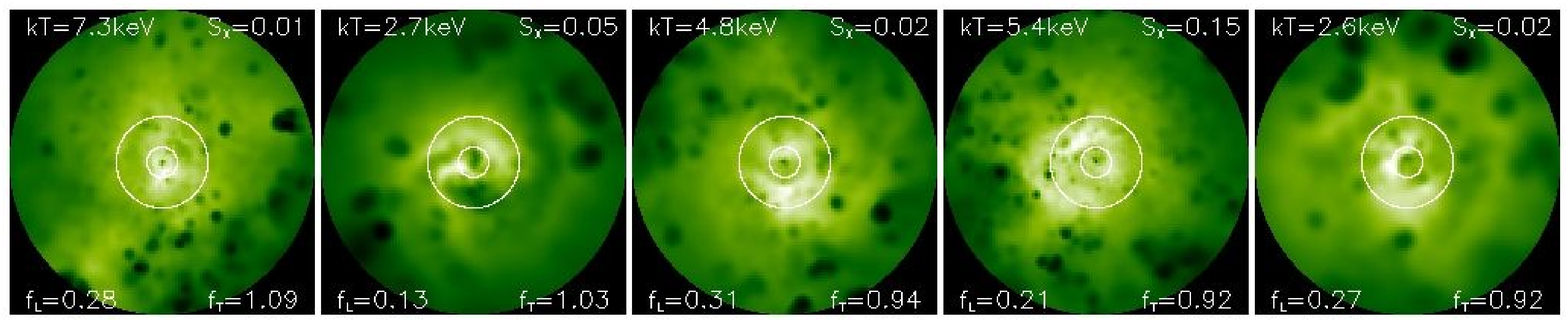}
\caption{Spectroscopic-like temperature maps of the 5 clusters with
the lowest (top panels) and highest (bottom panels) $f_{\rm T}$ values
at $z=0$. Circles mark the two inner radii (0.1 and 0.3 $R_{500}$)
where $f_{\rm T}$ is measured.  X-ray concentrations ($f_{\rm L}$) are
also given, which tend to be anti-correlated with $f_{\rm T}$ for
regular clusters.}
\label{fig:ftimages}
\end{figure*}

Two generic features of our simulated clusters at $z=0$ is that the
majority have cool cores (as shown in Fig.~\ref{fig:projtprof}) and
exhibit a large dispersion in core surface brightness
(Fig.~\ref{fig:sbprof}). At $z=1$, the clusters also tend to have cool
cores but a smaller range in core surface brightness is seen.  To
quantify this behaviour further, we define two simple core structure
parameters that are readily observable.

We first define three projected radii,
$[R_1,R_2,R_3]=[0.1,0.3,1.0]R_{500}$. The first approximately defines
the radius where the profile stops rising; the first and second
approximately define the (maximum) temperature plateau, and the third
is the outer radius of the cluster. The first parameter is then
\begin{equation}
f_{\rm T} = {T_{\rm sl}(<R_1) \over T_{\rm sl}(R_1\leq R \leq R_2)},
\end{equation}
which measures the ratio of the core to the maximum projected
spectroscopic-like temperature of the cluster. Clusters with the {\it
coolest} cores have the lowest $f_{\rm T}$ values. This can be seen
clearly in Fig.~\ref{fig:ftimages}, where temperature maps of the 5
clusters with the lowest and 5 with the highest $f_{\rm T}$ values are
shown. All but two clusters in our sample, including irregular systems,
have $f_{\rm T}<1$ because of their cool
cores; at $z=1$ the situation is similar, where only 6 clusters ($\sim
10$ per cent of the sample) have $f_{\rm T}>1$ (the median $f_{\rm T}$
increases gradually with redshift). We will return to this point in 
Section~\ref{sec:disc}.

The second structure parameter is 
\begin{equation}
f_{\rm L} = {L_{\rm bol}(<R_1) \over L_{\rm bol}(<R_3)},
\end{equation}
which measures the fraction of bolometric luminosity emanating from
the core; we label this the X-ray {\it concentration} of the cluster.
Clusters with the highest core surface brightness have the highest
$f_{\rm L}$ values.

\begin{figure}
\centering
\includegraphics[width=8cm]{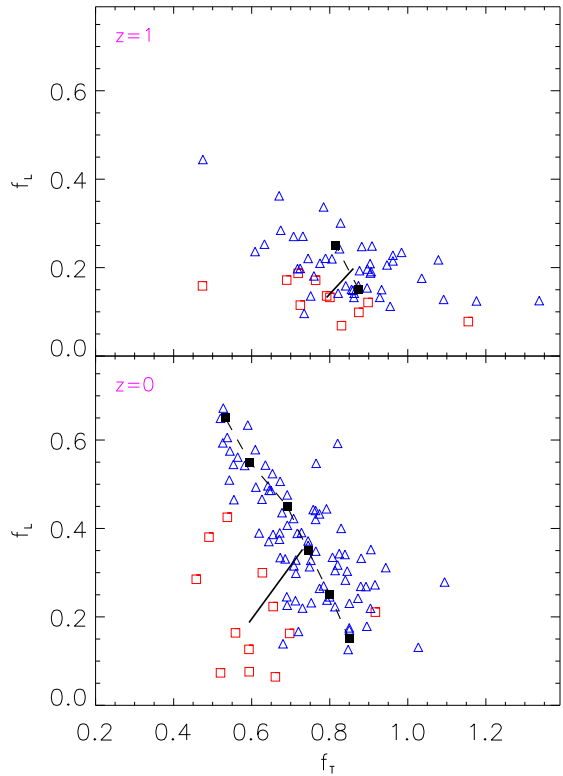}
\caption{X-ray concentration parameter, $f_{\rm L}$, versus
core-to-maximum temperature ratio, $f_{\rm T}$, for clusters at $z=1$
and $z=0$. Triangles are regular clusters
and squares irregular clusters, with their median ($f_{\rm T},f_{\rm
L}$) co-ordinates connected by a solid line. Solid squares, joined by
dashed lines, illustrate the median $f_{\rm T}$ at fixed intervals of
0.1 in $f_{\rm L}$, for regular clusters.}
\label{fig:ftfl}
\end{figure}

Like $S_{\rm X}$, values of $f_{\rm L}$ and $f_{\rm T}$ do not depend
strongly on temperature. Fig.~\ref{fig:ftfl} shows that the 
two quantities are anti-correlated for regular clusters, i.e. clusters 
with the highest X-ray concentrations have the coolest cores. As we
shall see, these systems tend to be older (less recent merger
activity), and thus the gas has had more time to settle down into a
regular state. Irregular clusters tend to have low $f_{\rm L}$
values as the subcluster boosts the overall luminosity
without affecting the core luminosity.

Examining the $f_{\rm L}$ distribution alone, there is a large spread in
values at $z=0$, varying from around $0.1-0.7$, but this reduces to
$\sim 0.1-0.4$ at $z=1$. As expected from Fig.~\ref{fig:sbprof}, we see
an absence of clusters with strongly-peaked X-ray emission at high
redshift (the median $f_{\rm L}$ decreases gradually with redshift).  
It is unlikely that this effect is due to poorer numerical
resolution at higher redshift, as nearly all our clusters have an inner
radius, $R_1$, that is larger than the physical softening radius,
$r_{\rm min}=56\hkpc/(1+z)$, at $z=0$ and $z=1$. Furthermore, the lack
of a strong dependence of $f_{\rm L}$ with temperature and the
presence of cool cores at all redshifts suggests that numerical
heating cannot be a major problem.

\subsection{Entropy profiles}
\label{subsec:entprof}

The combined effects of cooling and heating processes can be
effectively probed by measuring the entropy distribution of the ICM
and comparing this with the prediction of the gravitational-heating
model (see \citealt{Voit05} and references therein). Recently,
\citet{Voit05b} compared two independent sets of gravitational-heating
simulations and found that the outer entropy profiles were very
similar, $S \propto R^{1.2}$, close to the original prediction from
spherical accretion-shock models (e.g. \citealt{Tozzi01}). K2004a and
K2004b found that cooling and feedback (the same model used in this
paper) only slightly modified the outer slope of the entropy profile;
the main effect was an increase in the normalisation of the entropy at
all radii, as suggested by observations
(e.g. \citealt{Ponman03,Pratt03}).

\begin{figure}
\centering
\includegraphics[width=8cm]{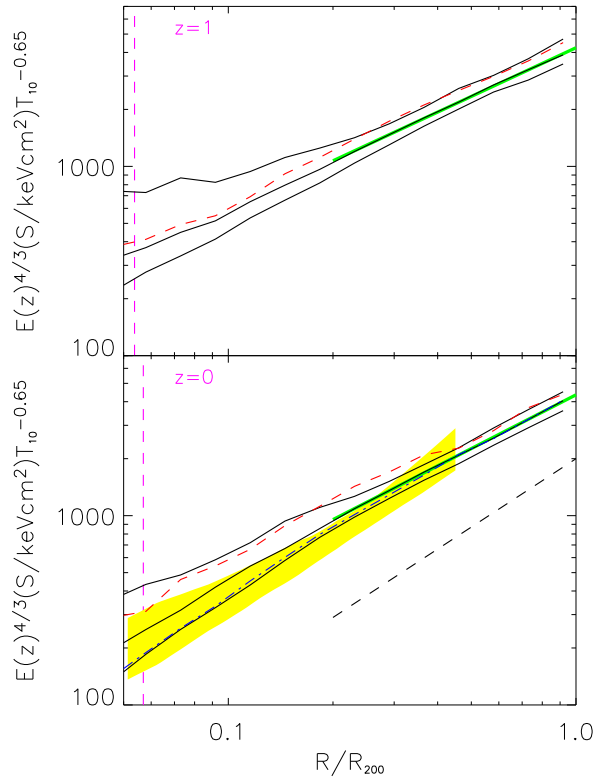}
\caption{Scaled entropy versus radius, in units of $R_{200}$, for
clusters at $z=1$ and $z=0$. The solid
curves are the median and 10/90 percentile profiles for regular
clusters, the dashed curve is the median profile for irregular
clusters and the dot-dashed curve for regular clusters with $f_{\rm
L}>0.45$.  The vertical dashed line illustrates the force resolution
of the simulation.  The large shaded region encloses the mean and
$1\sigma$ standard deviation entropy profile for an observed sample of
clusters by \citet{Pratt06}.  A least-squares fit to the simulated
data at $R>0.2R_{200}$ is illustrated by the thick solid line, flatter
than the prediction from gravitational-heating simulations, $S \propto
R^{1.2}$, shown with the lower dashed line \citep{Voit05b}.}
\label{fig:entprof}
\end{figure}

The CLEF simulation allows us to study entropy profiles for a much
larger sample of clusters than previously. In Paper~I, we showed that
the entropy--temperature relation at $z=0$ reproduced the observed
behaviour of excess entropy, both at small ($0.1R_{200}$) and large
($R_{500}$) radii.  In this paper, we focus on the scaled entropy
profiles at $z=0$ and $z=1$, shown in Fig.~\ref{fig:entprof}. We scale
the entropy profile of each cluster by $E^{4/3}(z)T_{10}^{-0.65}$; the
first factor reflects the predicted redshift scaling from
gravitational heating, while the second approximately represents the
scaling with temperature at fixed radius/overdensity, modified by
non-gravitational processes. As will be explained below, we define
$T_{10}$ to be the projected spectroscopic-like temperature in $10 {\rm
keV}/k$ units, measured between $R_1$ and $R_2$ (i.e. the temperature
plateau, see subsection~\ref{subsec:structparam}).

The solid curves illustrate the median and 10/90 percentile profiles
for regular clusters at each redshift. At both redshifts the profiles
are close to power-law outside the core ($R \sim (0.2-1)R_{200}$);
fitting a straight line to the profile in this region yields $S
\propto R^{0.9}$ at both redshifts, as found in previous papers
(K2004a,b). Note that at $R_{500}$, the normalisation of the scaled
profile is very similar at $z=0$ and $z=1$, thus the entropy at large
radii scales with redshift as predicted from gravitational-heating
models.

We also plot the median profile of irregular clusters (shown as the
dashed curve) and at $z=0$, clusters with the highest X-ray 
concentrations ($f_{\rm L}>0.45$). Irregular clusters
tend to have higher entropy profiles than the regular clusters at all
radii and at both redshifts. This temporary elevation in entropy
reflects the shock-heating processes associated with the merger. 
Conversely, clusters with the highest X-ray concentrations have the
lowest entropy profiles, reflecting the fact that they have the highest 
cooling rate. We also note that the
profile for these systems resembles a broken power law, similar to
that observed by \citet{Finoguenov05} in their REFLEX-DXL clusters
($z\sim 0.3$).

We compare the profiles at $z=0$ to the recent {\it XMM-Newton} data
studied by \citet{Pratt06} (the mean plus/minus $1\sigma$ values are
shown as the shaded region). \citet{Pratt06} use a global mean
temperature, measured between $0.1-0.5R_{200}$, for the entropy
scaling. We note, however, that this effectively measures their
temperature plateau as they see no significant evidence of a decline
at large radii \citep{Arnaud05}.  The simulated (regular clusters) and
observed distributions are similar, although the simulated profile is
slightly high. We note, however, that our regular, concentrated
clusters fit the observational profile very well; the observed sample
is probably biased to systems of this type as they are intrinsically
brighter systems and are thus easier to observe.

\subsection{Mass estimates}
\label{subsec:mest}

Finally in this Section, we briefly investigate the validity of
hydrostatic equilibrium in the simulated clusters at all redshifts,
used to estimate cluster masses from X-ray data
\begin{equation}
M_{\rm est}(<r) = -{r kT(r) \over G\mu m_{\rm H}} \, 
\left[ {d \ln \rho \over d \ln r} + {d\ln T \over d\ln r} \right],
\label{eqn:mest}
\end{equation}
where $\rho(r)$ and $T(r)$ are the 3D density and spectroscopic-like
temperature profiles respectively. Various approximations to
equation~(\ref{eqn:mest}) have been used previously in the literature,
when little or no spatial information was available for the
temperature distribution in clusters. Newer, high-quality observations
with {\it XMM-Newton} and {\it Chandra} have overcome this problem
(e.g.~\citealt{Arnaud05,Vikhlinin05b}), and so we assume in our study
that the gas density and temperature profiles can be accurately
recovered from the X-ray data out to $R_{500}$.  (A detailed study of
obtaining such profiles from mock X-ray data is left to future work,
but see \citealt{Rasia06}.)

\begin{figure}
\includegraphics[width=8cm]{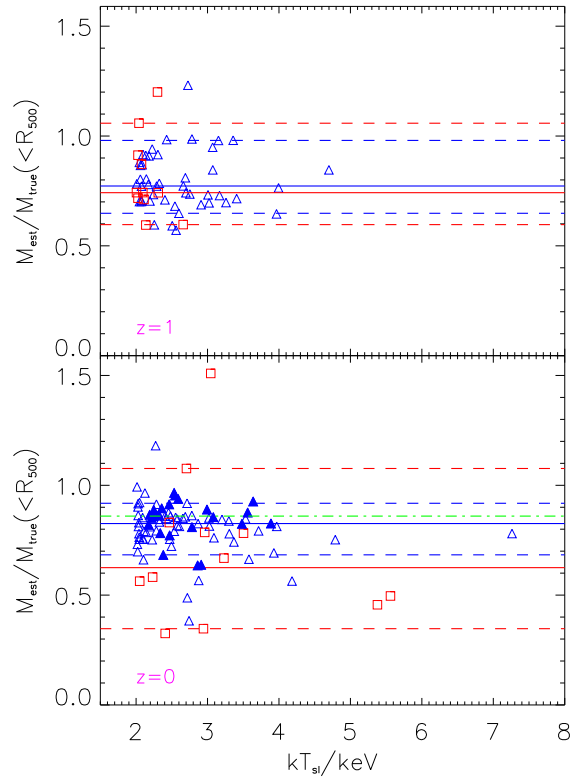}
\caption{Ratio of estimated to true masses at $R_{500}$
versus $T_{\rm sl}$ for clusters at $z=1$ and $z=0$. 
Squares are irregular clusters, triangles are regular
clusters and filled triangles regular clusters with the
highest X-ray concentrations, $f_{\rm L}>0.45$. Solid (dashed) 
lines are median (10/90 percentile) ratios for regular and 
irregular clusters. The dot-dashed line is the
median ratio for regular clusters with $f_{\rm L}>0.45$.}
\label{fig:mesttsl}
\end{figure}

In Fig.~\ref{fig:mesttsl}, we plot estimated to true mass ratios
at $R_{500}$ for our clusters at $z=1$ and
$z=0$. For internal consistency, estimated masses are those
at the estimated $R_{500}$, which is typically 5 per cent smaller
than the true $R_{500}$. 

The median estimated mass is around 80 per cent of the true mass
at both redshifts for regular clusters, with 10-20 per cent scatter.
Irregular clusters tend to have slightly poorer mass estimates on
average, but the scatter is also larger. Clusters with the highest
X-ray concentrations perform slightly better than the regular clusters
as a whole.

Our median mass ratio is lower than found by K2004b at $z=0$, who
found that the average estimated mass was only 5 per cent or so lower
than the true mass at $R_{500}$, with the small discrepancy being due
to turbulent motions (see also \citealt{Evrard96,Rasia04}). The reason
for the difference is twofold. Firstly, K2004b used a mass-weighted
temperature profile, where we use the more realistic
spectroscopic-like temperature profile. This reduces mass estimates by
10 per cent or so, similar to what was found by \citet{Rasia06}, when
assuming a low X-ray background in their analysis.  A further 5 per
cent reduction comes from using the estimated $R_{500}$ rather than
the true value, which also increases the scatter. While we are
therefore not comparing true and estimated masses at the same radius
here, we are demonstrating what the overall effect will be on the
normalisation of the mass--temperature relation, as will be
investigated in the next section.

\section{Cluster Scaling Relations}
\label{sec:scale}

We now put together the results from previous sections to try and
understand the properties of cluster scaling relations in our
simulation. We consider the two most important X-ray scaling relations
in this paper: mass versus temperature ($M-T$) and luminosity versus
temperature ($L-T$), with all quantities computed within $R_{500}$.

Scaling relations are defined using the conventional form
\begin{equation}
Y = Y_0 \, (X/X_0)^{\alpha} \, (1+z)^{\beta},
\label{eqn:srel}
\end{equation}
where $Y_0$ is the normalisation at $X=X_0$ and $z=0$ (for all
relations, $X_0=5{\rm keV}$ ; for the $M-T$ relation,
$Y_0=M/10^{14}\hMsol$ and for the $L-T$ relation,
$Y_0=L/10^{44}h^{-2}{\rm erg s}^{-1}$); $\alpha$ is the slope
and $\beta$ the parameter used to
describe the redshift dependence of the normalisation.  Scatter in the
relations is measured at each redshift as
\begin{equation}
\sigma_{\log(Y)} = \sqrt{{1 \over N} \, \sum_i \, (\log(Y_i/Y))^2}
\label{eqn:scatter}
\end{equation}
i.e. the r.m.s. deviation of $\log(Y)$ from the mean relation, where
$Y_i$ are individual data points. We then parameterise any redshift
dependence of the scatter using a least-squares fit to
$\sigma_{\log(Y)}$
\begin{equation}
\left< \sigma_{\log(Y)}\right> (z) = \sigma_0 + \sigma_1 \log(1+z).
\label{eqn:scatterz}
\end{equation}

\subsection{Results at $z=0$}

\begin{table}
\caption{Best-fit parameter values (and $1\sigma$ errors) for scaling
relations at $z=0$. Column~1 gives the sample used in the fit; column 2
the best-fit normalisation; column 3 the best-fit slope and column 4
the logarithmic scatter.}
\begin{center}
\begin{tabular}{lccc}

\hline
Sample & $Y_0$ & $\alpha$ & $\sigma_{\log(Y)}$\\
\hline

\multicolumn{4}{l}{$M_{500}-T_{\rm dyn}$}\\
All  & $3.44 \pm 0.07$ & $1.38 \pm 0.03$ & $0.05$\\
Reg  & $3.46 \pm 0.08$ & $1.38 \pm 0.04$ & $0.05$\\
High & $3.53 \pm 0.11$ & $1.38 \pm 0.07$ & $0.03$\\
Low  & $3.40 \pm 0.11$ & $1.36 \pm 0.05$ & $0.05$\\

\hline

\multicolumn{4}{l}{$M_{500}-T_{\rm gas}$}\\
All  & $4.08 \pm 0.06$ & $1.69 \pm 0.03$ & $0.03$\\
Reg  & $4.05 \pm 0.06$ & $1.67 \pm 0.03$ & $0.03$\\
High & $3.95 \pm 0.12$ & $1.59 \pm 0.06$ & $0.03$\\
Low  & $4.07 \pm 0.08$ & $1.69 \pm 0.03$ & $0.03$\\

\hline

\multicolumn{4}{l}{$M_{500}-T_{\rm sl}$}\\
All  & $5.51 \pm 0.32$ & $1.81 \pm 0.08$ & $0.08$\\
Reg  & $4.93 \pm 0.29$ & $1.69 \pm 0.08$ & $0.08$\\
High & $7.03 \pm 0.78$ & $1.96 \pm 0.16$ & $0.06$\\
Low  & $4.37 \pm 0.21$ & $1.61 \pm 0.07$ & $0.06$\\

\hline

\multicolumn{4}{l}{$M_{500}-T_{\rm sl}^{50}$}\\
All  & $4.47 \pm 0.19$ & $1.76 \pm 0.07$ & $0.08$\\
Reg  & $4.02 \pm 0.12$ & $1.64 \pm 0.05$ & $0.05$\\
High & $3.94 \pm 0.13$ & $1.62 \pm 0.07$ & $0.03$\\
Low  & $4.08 \pm 0.17$ & $1.66 \pm 0.06$ & $0.05$\\

\hline

\multicolumn{4}{l}{$M_{500}^{\rm est}-T_{\rm sl}^{50}$}\\
All  & $3.28 \pm 0.15$ & $1.64 \pm 0.08$ & $0.09$\\
Reg  & $3.11 \pm 0.11$ & $1.57 \pm 0.06$ & $0.06$\\
High & $3.22 \pm 0.15$ & $1.56 \pm 0.10$ & $0.04$\\
Low  & $2.98 \pm 0.15$ & $1.53 \pm 0.08$ & $0.06$\\

\hline


\multicolumn{4}{l}{$L_{\rm bol}-T_{\rm sl}$}\\
All  & $6.0  \pm 1.1$ & $3.08 \pm 0.26$ & 0.27\\
Reg  & $5.8  \pm 1.3$ & $3.04 \pm 0.30$ & 0.27\\
High & $19.0 \pm 6.5$ & $3.61 \pm 0.45$ & 0.16\\
Low  & $3.6  \pm 0.5$ & $2.81 \pm 0.19$ & 0.16\\

\hline

\multicolumn{4}{l}{$L_{\rm bol}^{50}-T_{\rm sl}^{50}$}\\
All  & $4.3 \pm 0.3$ & $3.45 \pm 0.13$ & 0.14\\
Reg  & $3.9 \pm 0.3$ & $3.37 \pm 0.12$ & 0.12\\
High & $4.7 \pm 0.3$ & $3.16 \pm 0.12$ & 0.05\\
Low  & $3.1 \pm 0.3$ & $3.13 \pm 0.14$ & 0.11\\

\hline

\end{tabular}
\end{center}
\label{tab:srelz0}
\end{table}

Our results for $z=0$ clusters are summarised in
Table~\ref{tab:srelz0}.  Column~1 lists the sample used when fitting
the data. Here, we consider all 95 clusters in our temperature-limited
sample (labelled {\it All}), the 83 regular clusters (i.e. those with
$S_{\rm X}\leq 0.1$; labelled {\it Reg}), the 23 regular clusters with
the most prominent core emission ($f_{\rm L}>0.45)$; denoted {\it
High}) and the 60 remaining regular clusters (denoted {\it Low}).
Column~2 lists the best-fit normalisation, $Y_0$; column~3 the slope
of the relation, $\alpha$; and column~4 the scatter in the relation,
$\sigma_{\log(Y)}$.  We now discuss each relation in turn.

\subsubsection{$M_{500}-T$ relation}

\begin{figure}
\includegraphics[width=8cm]{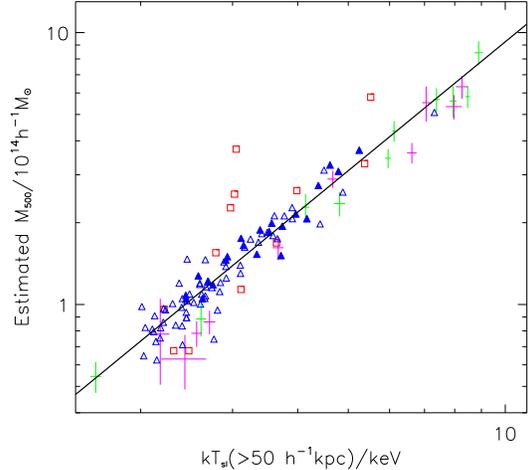}
\caption{Estimated scaled-mass at $R_{500}$ versus spectroscopic-like
temperature outside the core at $z=0$. 
Squares are irregular clusters, triangles are regular
clusters and filled triangles regular clusters with the
highest X-ray concentrations, $f_{\rm L}>0.45$.
The solid line in each panel is a best fit to
regular clusters. Crosses are data-points from \citet{Arnaud05} and
\citet{Vikhlinin05b}.}
\label{fig:mt}
\end{figure}

\begin{figure*}
\includegraphics[width=8cm]{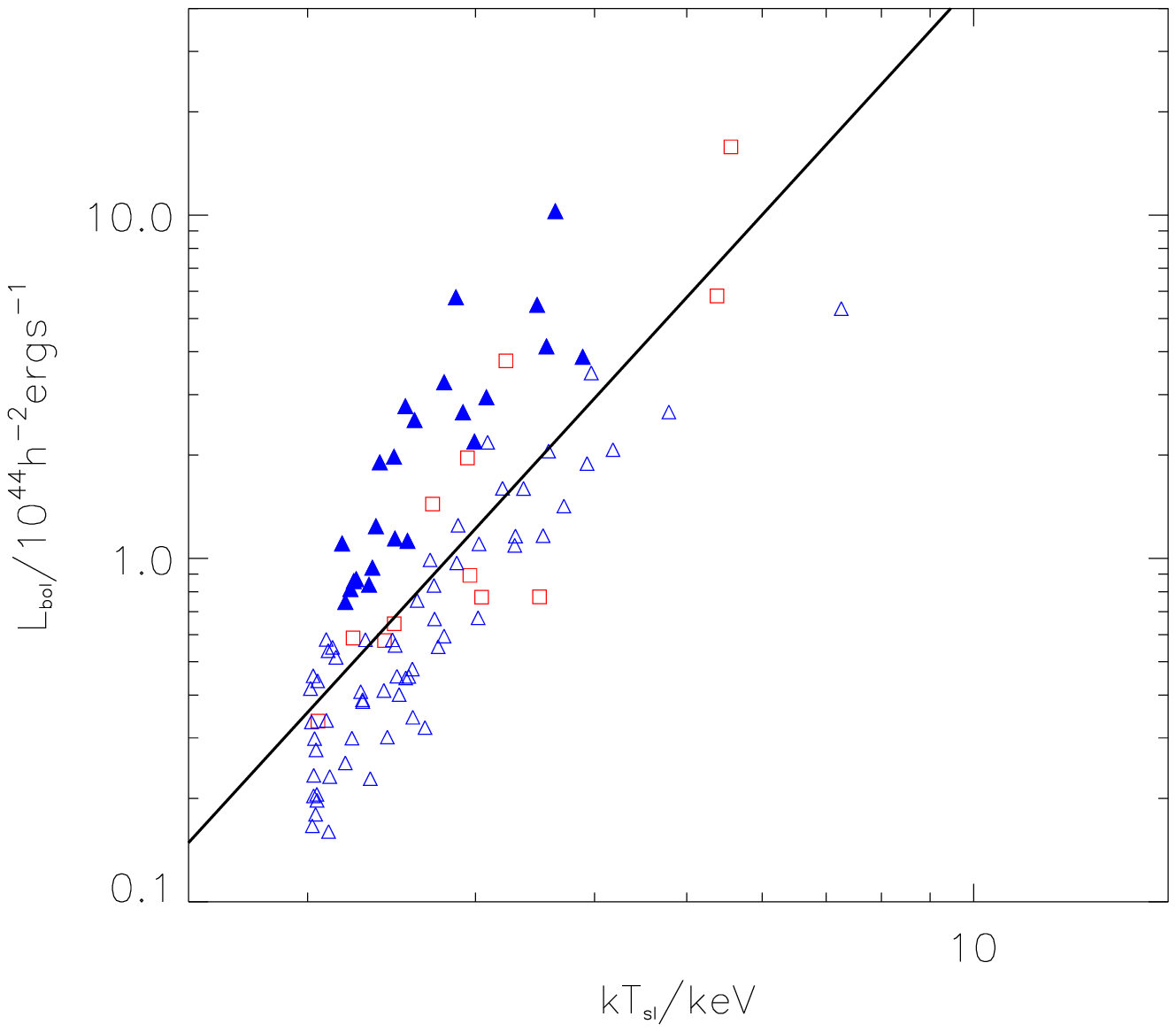}
\includegraphics[width=8cm]{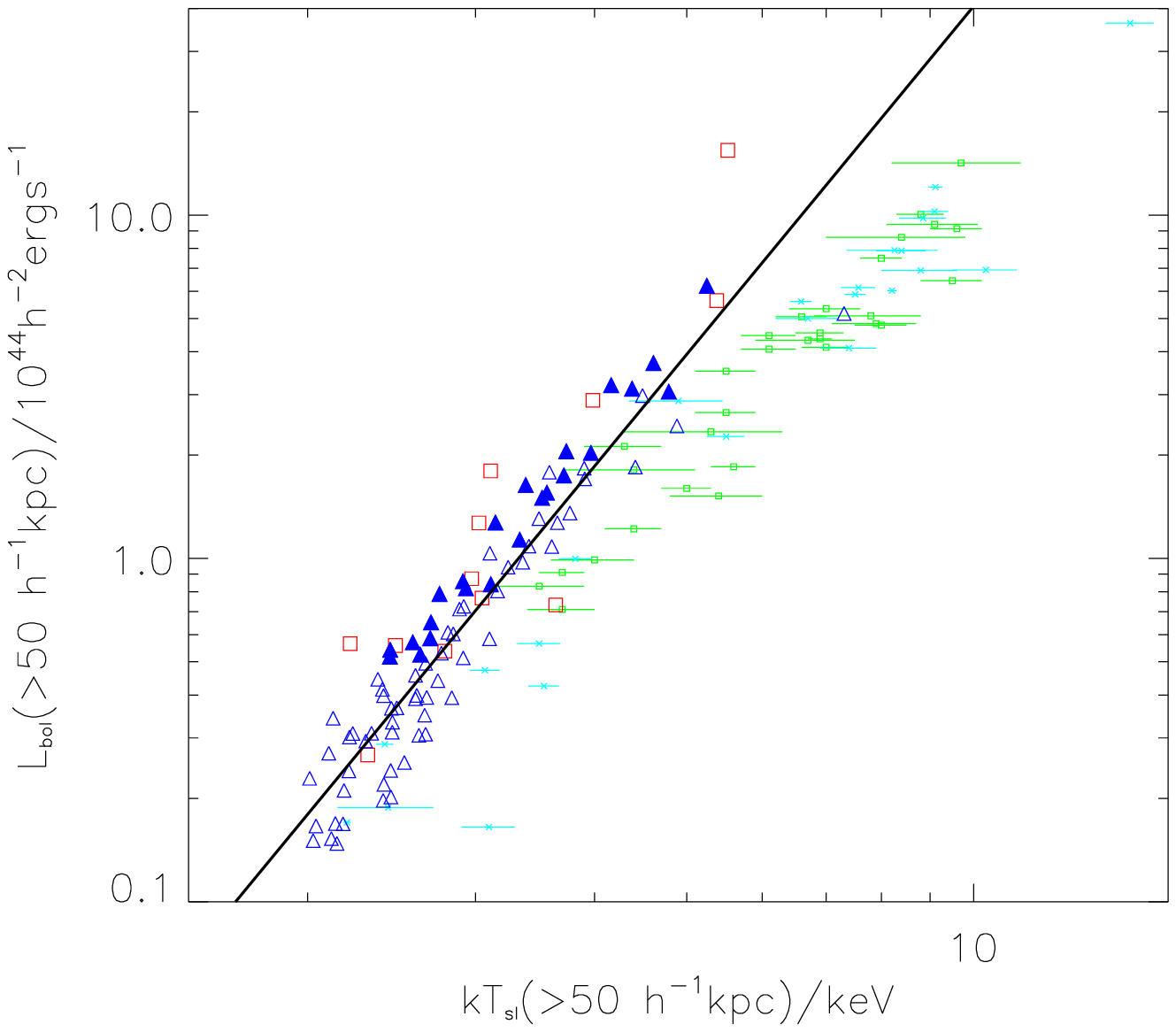}
\caption{Bolometric luminosity versus spectroscopic-like temperature,
for all emission (left panel) and emission outside the core (right
panel). 
Squares are irregular clusters, triangles are regular
clusters and filled triangles regular clusters with the
highest X-ray concentrations, $f_{\rm L}>0.45$.
The solid line in each
panel is a best fit to regular clusters. Data points with error bars
are observed values from \citet{Markevitch98} and \citet{Arnaud99}.}
\label{fig:lt}
\end{figure*}

We first study the $M-T$ relation at $z=0$. In Paper~I we presented
results for the hot gas mass-weighted temperature within $R_{2500}$,
$T_{\rm gas}$, and showed that the relation was in good agreement with
the {\it Chandra} results of \citet{Allen01}. Here we discuss the
$M$--$T$ relation at $R_{500}$ as we expect it to be less susceptible
to cooling and heating effects associated with the cluster core.
While measuring the relation at $R_{500}$ is observationally
challenging, even with {\it XMM-Newton} and {\it Chandra}, recent
attempts have been performed for a small sample of clusters with a
reasonable range in temperature (\citealt{Arnaud05}, hereafter APP05;
\citealt{Vikhlinin05b}, hereafter VKF06).  We will eventually compare
our results at $z=0$ to these observations, but first study how our
definition of temperature and mass affects the details of the
relation.

We initially consider the relation between the true total mass of a
cluster, $M_{500}$, and its dynamical temperature, $T_{\rm dyn}$,
where the latter was defined in equation~(\ref{eqn:tdyn}). This relation
should most faithfully represent the scaling expected from
gravitational-heating models ($\alpha=1.5$) but as listed in
Table~\ref{tab:srelz0}, the measured slope is slightly shallower than
this ($\alpha \sim 1.4$). As discussed in \citet{Muanwong06}, the
deviation in slope is consistent with the variation in halo
concentration with cluster mass (i.e. even the dark matter is not
perfectly self-similar). Note that the sub-samples give almost
identical results to the overall sample, although the {\it High}
sub-sample exhibits less scatter.

We next consider the hot gas mass-weighted temperature, $T_{\rm
gas}$. The slope of the relation steepens to $\alpha \sim 1.7$; as
discussed in Section~\ref{subsec:clustemp}, this is due to the
combined effects of heating and cooling. Strikingly, the scatter in
this relation is very small ($\sigma_{\log(T)}=0.03$). Again, no
significant change in the relation is observed when the cluster
sub-samples are considered.

When the X-ray temperature, $T_{\rm sl}$, is used, both the
normalisation and scatter increase, with the irregular and {\it High}
clusters lying above the mean relation (i.e. they are colder than
average). This is because cool, dense gas in the core and in
substructures throughout the cluster (see Fig.~\ref{fig:sltimages}) is
weighted more heavily than before, and there is a large variation in
the cool gas distribution from cluster to cluster (see also
\citealt{Muanwong06}; \citealt{OHara06}). As can be seen in
Fig.~\ref{fig:ftfl}, the irregular and the {\it High} clusters have
the lowest $f_{\rm T}$ values.

We also present results for the spectroscopic-like temperature when
particles from within the inner $50 \hkpc$ core are excluded (denoted
$T_{\rm sl}^{50}$), which reduces the scatter in the {\it Reg}
clusters from 0.08 to 0.05. The {\it High} and {\it Low} relations
are now consistent with the overall {\it Reg} relation, although 
the irregular clusters still lie above the relation as a 
second cool core is still present.

Finally, we replace the actual mass with the mass estimated under the
assumption of hydrostatic equilibrium (denoted $M_{500}^{\rm est}$),
as defined in Section~\ref{subsec:mest}. 
Fig.~\ref{fig:mt} illustrates the result, in comparison to the data
from APP05 and VKF06.  The relation for the {\it Reg} subsample
provides the closest match to the observational data. The main effect
of using the estimated mass is to reduce the normalisation by $\sim
20$ per cent. Although the two observational
samples are similar, our {\it Reg} relation is closest to the best-fit
results of VKF06; the slope and scatter are almost identical (VKF06
find $\alpha=1.58$ and $\sigma_{\log(M)}\sim 0.06$) and the
normalisation differs by 10 per cent or so 
(VKF06 find $Y_0=2.89 \pm 0.15$).
Given the variations between parameters considered in this study, this
is quite a good match, but serves to point out that a precision
measurement of the $M-T$ relation is non-trivial and must include
several physical effects; our present study is by no means exhaustive
(see \citealt{Rasia06}).

\subsubsection{$L-T$ relation}

To study the luminosity--temperature relation, we compute bolometric
luminosities, $L_{\rm bol}$, for all emission within $R_{500}$ (where
more than 90 per cent of the cluster emission comes from). We also
compute luminosities outside the core (denoted $L_{\rm bol}^{50}$),
again by excluding all hot gas particles from within $50\hkpc$ from
the cluster's centre.

Fig.~\ref{fig:lt} illustrates
luminosity--temperature relations at $z=0$.  In the left panel we show
results for total luminosities and spectroscopic-like temperatures,
and in the right panel, for luminosities and temperatures outside the
$50\hkpc$ core. Best-fit parameters for the various cluster samples at
$z=0$ are also given in Table~\ref{tab:srelz0}.

When all emission is included, the $L-T$ relation at $z=0$ has a large
amount of scatter. Comparing the relation for regular clusters with
high $f_{\rm L}$ values to those with low $f_{\rm L}$ values, we see
that the two subpopulations are widely separated in the $L-T$
plane. The scatter thus reflects the strength of the core emission, as
shown observationally by \citet{Fabian94}. We discuss this further in
Section~\ref{sec:disc}.

When the $50\hkpc$ core
emission is excised, the scatter in the relation reduces
substantially, from 0.27 to 0.14, with all samples then having very
similar properties. We also note that irregular clusters do not lie
systematically off the $L-T$ relation, in agreement with \citet{Rowley04},
who analysed a simulation with radiative cooling but no feedback.

We compare our excised-core results with the observational data of
\citet{Markevitch98} and \citet{Arnaud99}; the former also excised
emission from the inner $50\hkpc$ and the latter selected {\it
non-cooling-flow} clusters.  Although our clusters do not cover the
same dynamic range as the observations, we note that our $L-T$ relation 
has a normalisation that is too high (see also Paper~I). This suggests
that that cluster temperatures in general are too low (note that 
higher temperatures may not significantly affect the normalisation of 
the $M-T$ relation, as the estimated mass depends linearly on $T$). The
slope of the relation for all clusters, $\alpha=3.5$, is 
steeper than the observations ($\alpha \sim 2.6-2.9$).  As stated in
Paper~I, the slope varies systematically with temperature, such that
higher-mass clusters have lower values. The lack of hot clusters in
our sample biases our result to higher values. Significantly larger
volumes are still required to capture the rich clusters, to get a more
accurate (average) slope for the cluster population.

\subsection{Evolution of scaling relations}

\begin{figure*}
\includegraphics[width=8.5cm]{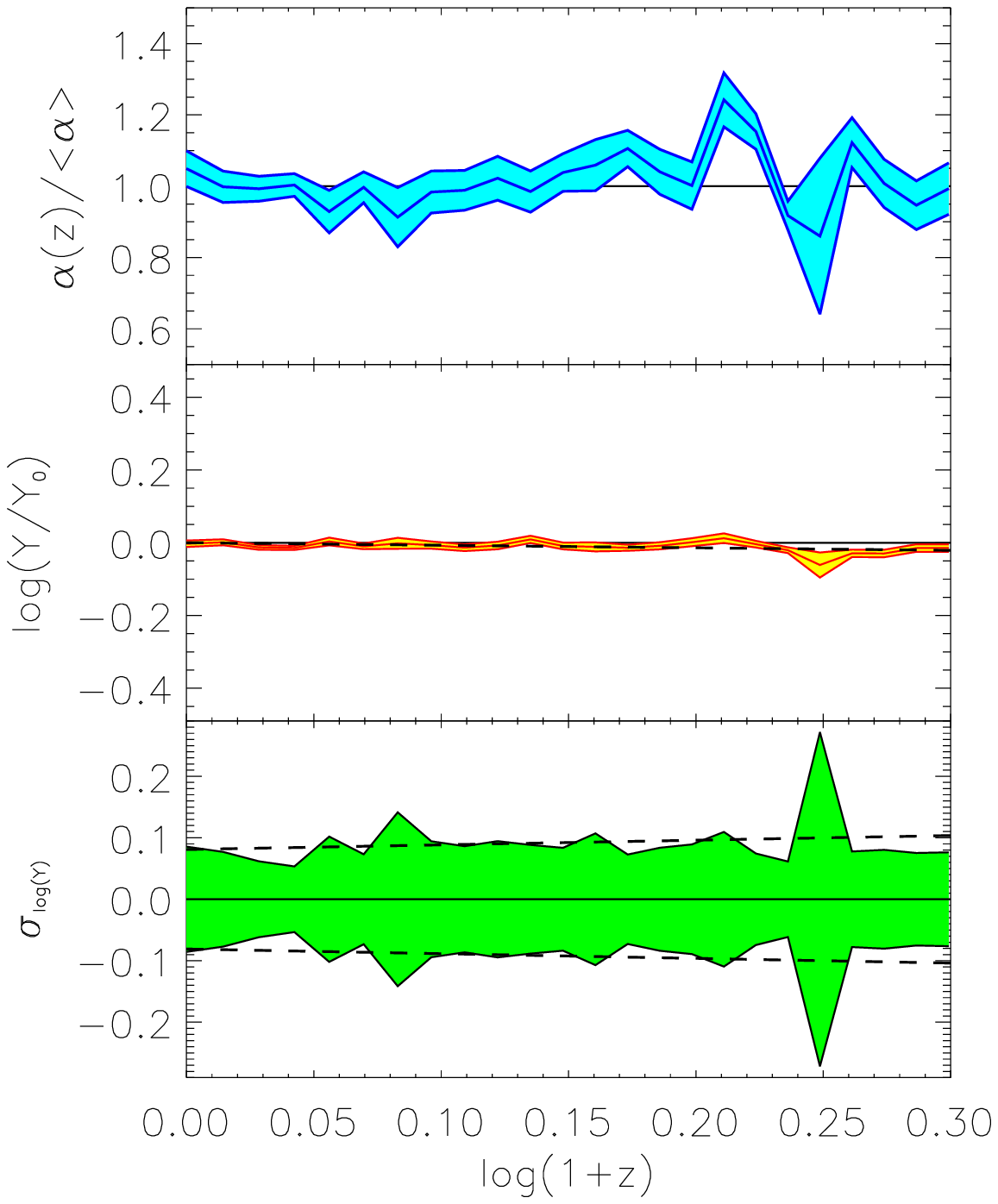}
\includegraphics[width=8.5cm]{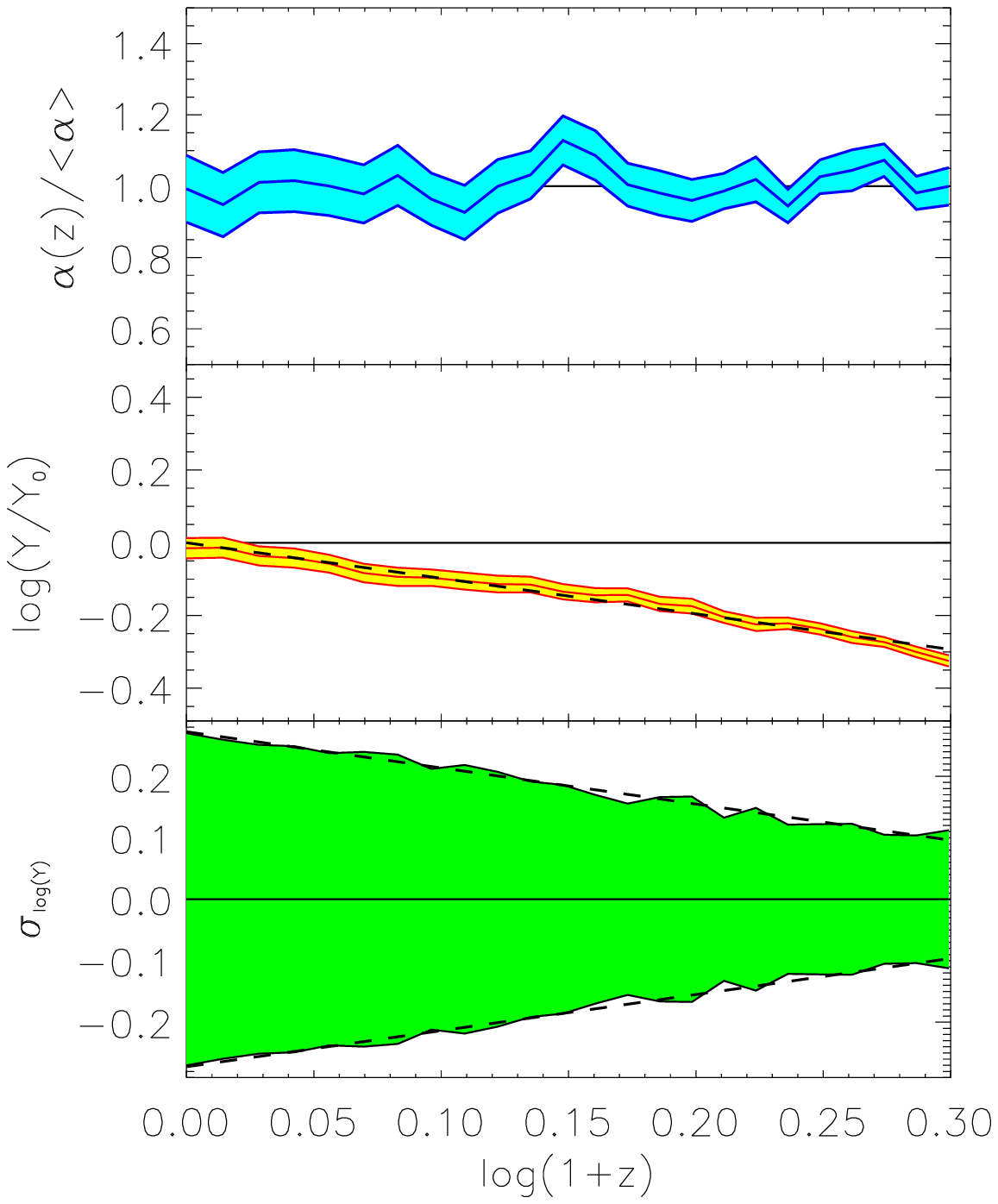}
\caption{Redshift dependence of the slope, normalisation and scatter
of the $M_{500}^{\rm est}-T_{\rm sl}^{50}$ (left panel) and $L_{\rm
bol}-T_{\rm sl}$ (right panel) scaling relations. The band in the top
panel illustrates the mean slope plus the standard deviation at each
redshift, all normalised to the median value over all redshifts
between $z=0-1$. The band in the middle panel illustrates the mean and
standard deviation of the normalisation at each redshift (assuming a
fixed slope, at the median value). The best-fit straight line to the mean
data is also plotted as a dashed line and is used to normalise the
data at $z=0$. The band in the bottom panel illustrates the
logarithmic scatter in the scaling relations at each redshift (values
are reflected about the x-axis to give an idea of the full size of
scatter in the relation), with the best-fit straight line given by the
dashed line.}
\label{fig:srelz}
\end{figure*}

We now study how the $M-T$ and $L-T$ relations evolve with
redshift. We first measure the slope, normalisation and scatter of the
relations at each redshift between $z=0$ and $z=1$. The
gravitational-heating model predicts the slope to be constant with
redshift. For the $M-T$ relations this is generally true; although the
variation can be quite noisy, there is no evidence for a systematic
change in the slope, $\alpha$, with redshift (e.g. see the top-left
panel in Fig.~\ref{fig:srelz} for how the slope changes with redshift
in the $M_{500}^{\rm est}-T_{\rm sl}^{50}$ relation). For the $L-T$
relation (all emission), the slope increases with redshift when all clusters
are considered. This is because the few hottest clusters have anomalously
high temperatures for their luminosity at low redshift, causing a decrease in 
slope since they carry a lot of weight. At higher redshift the effect diminishes
as the clusters move back towards the mean relation. We circumvent this problem 
by restricting our fit to the $L-T$ relation to clusters with $2<kT<5$ keV at each 
redshift; as can be seen in the top-right panel of Fig.~\ref{fig:srelz}, the slope of the 
$L_{\rm bol}-T_{\rm sl}$ relation is now approximately constant. For all relations, we 
fix $\alpha$ to its median value between $z=0-1$.

With $\alpha$ determined, we then fit equation~(\ref{eqn:srel}) to the
normalisation data to determine $Y_0$ and $\beta$. (Note this may
cause $Y_0$ to change slightly from the exact $z=0$ value.) The
scatter is also determined at each redshift
(equation~\ref{eqn:scatter}) and fit with equation~(\ref{eqn:scatterz}).

\begin{table*}
\caption{Best-fit parameter values (and $1\sigma$ errors) for
evolution of scaling relations from $z=0-1$. Column 1 gives the median
slope used for the fit; columns 2 \& 3 the best-fit
normalisation and evolution parameters; and columns 4 \& 5 the
best-fit scatter parameters.}
\begin{center}
\begin{tabular}{ccccc}

\hline 
$\left< \alpha \right>$ & $Y_0$ & $\beta$ & $\sigma_0$ & $\sigma_1$\\ 
\hline

\multicolumn{3}{l}{$E(z)M_{500}-T_{\rm dyn}$}\\
$1.40$ & $3.48 \pm 0.02$  & $-0.07 \pm 0.01$ & $0.05 \pm 0.002$ &
$-0.02  \pm 0.01$\\ 

\hline

\multicolumn{3}{l}{$E(z)M_{500}-T_{\rm gas}$}\\
$1.68$ & $4.08 \pm 0.02$  & $-0.12 \pm 0.01$ & $0.04 \pm 0.004$ &
$0.03  \pm 0.02$\\ 

\hline 

\multicolumn{2}{l}{$E(z)M_{500}-T_{\rm sl}$}\\
$1.77$ & $5.21 \pm 0.04$  & $-0.13 \pm 0.02$ & $0.08 \pm 0.003$ &
$-0.05 \pm 0.01$\\ 

\hline

\multicolumn{2}{l}{$E(z)M_{500}-T_{\rm sl}^{50}$}\\
$1.67$ & $4.24 \pm 0.03$  & $0.04 \pm 0.02$ & $0.07 \pm 0.003$ &
$-0.03 \pm 0.02$\\ 

\hline

\multicolumn{2}{l}{$E(z)M_{500}^{\rm est}-T_{\rm sl}^{50}$}\\
$1.56$ & $3.17 \pm 0.04$  & $-0.07 \pm 0.03$ & $0.08 \pm 0.02$ &
$0.08 \pm 0.10$\\ 

\hline

\multicolumn{2}{l}{$E^{-1}(z)L_{\rm bol}-T_{\rm sl}$}\\
$3.36$ & $7.36 \pm 0.09$  & $-0.98 \pm 0.03$ & $0.27 \pm 0.003$ &
$-0.59 \pm 0.02$\\ 

\hline 

\multicolumn{2}{l}{$E^{-1}(z)L_{\rm bol}^{50}-T_{\rm sl}^{50}$}\\
$3.41$ & $4.53 \pm 0.07$  & $-0.61 \pm 0.04$ & $0.14 \pm 0.004$ &
$-0.23 \pm 0.02$\\ 

\hline

\end{tabular}
\end{center}
\label{tab:srelz}
\end{table*}

Table~\ref{tab:srelz} gives best-fit parameters for our generalised
scaling relations when applied to all clusters at each redshift.
For the $E(z)M-T$ relations, we see a lack of evolution relative
to the simple scalings predicted from gravitational heating, 
with $|\beta| \approxlt 0.15$. The scatter also changes very little
with redshift, with $|\sigma_1|<0.1$ in all cases. This lack of evolution 
in normalisation and scatter is illustrated more clearly for the 
$E(z)M_{500}^{\rm est}-T_{\rm sl}^{50}$ relation in the left panels 
of Fig.~\ref{fig:srelz}.

The evolution of the $E^{-1}(z)L-T$ relation is also presented in
Fig.~\ref{fig:srelz} (see also Table~\ref{tab:srelz}). Contrary to the
$M-T$ relation, this relation evolves negatively with redshift, with
$\beta \sim -1$. Note the amount of evolution at $z=1$ is comparable to the
intrinsic scatter in the relation at $z=0$. What is striking from the
figure, however, is the evolution of the scatter with redshift:
$\sigma_{\log(L)}$ at $z=1$ is almost a factor of 3 lower than at
$z=0$. As was found in subsection~\ref{subsec:structparam}, the
dispersion in X-ray concentration decreases with redshift, such that
at high redshift, clusters with strong cooling cores are absent. This
is reflected here as a reduction in the scatter of the $L-T$
relation. When the core is excised, the scatter is reduced at all
redshifts and also evolves less. The normalisation also evolves less
with redshift, demonstrating that some (but not all) of the deviation
from the gravitational-heating case is due to processes occurring
within the inner core.  Furthermore, since we know that the
$E(z)M_{500}-T_{\rm sl}^{50}$ relation evolves very weakly with
redshift, negative evolution in the $E^{-1}(z)L_{\rm bol}^{50}-T_{\rm
sl}^{50}$ relation is almost entirely due to a deficit in luminosity,
again as seen in the entropy and surface-brightness profiles.

A similar study was performed by \citet{Ettori04}, using the same
simulation as \citet{Borgani04}. Although they used a different model
for star formation and feedback than used here, they obtained very
similar results for the evolution of the $E(z)M-T$ and $E^{-1}(z)L-T$
relations; using our notation, they found $\beta=-0.2$ and
$\beta=-0.8$ respectively. On the other hand, \citet{Muanwong06}
compared a simulation similar to (but smaller than) the CLEF
simulation, with a simulation with radiative cooling only and with a
simulation with cooling and preheating. They found that the evolution
of the $L-T$ relation varied enormously between the models. Their
conclusion was that the amount of evolution depended on the nature of
non-gravitational processes.  We can thus conclude, at this point, that
no general consensus has emerged from numerical simulations as to
what the expected evolution of cluster scaling relations will be, once
sufficiently-large samples of high-redshift clusters exist. Of vital
importance, from the simulation side, will be to produce cluster
catalogues that are well matched to the observations; in particular,
the deficit of high-temperature systems in most studies to date needs
to be addressed.

\section{Discussion}
\label{sec:disc}

Perhaps the most interesting result in this paper is that our
simulation predicts a large scatter in the luminosity--temperature
relation at low redshift, as observed, however this scatter decreases
with redshift due to the lack of systems with high X-ray concentrations
at $z \sim 1$. Here, we discuss this issue in more detail and investigate 
further the differences between clusters with high and low X-ray
concentrations, and cool and warm cores. 

\subsection{Mass deposition rates}
\label{subsec:mdotx}

Observed samples of (generally low-redshift) clusters are historically
split into {\it cooling-flow} and {\it non-cooling-flow} systems
(e.g. \citealt{Fabian94}), with the former having higher mass
deposition rates, usually estimated from their core luminosity and
temperature
\begin{equation} 
\dot{M}_{\rm X} = {2 \over 5} \, {\mu m_{\rm H} L \over kT}.
\label{eqn:mdotx}
\end{equation}
X-ray spectroscopy of cluster cores has revealed that significantly
less gas in high $\dot{M}_{\rm X}$ clusters is actually cooling down
to temperatures significantly below the mean temperature of the
cluster. This lack of cold gas is likely attributed to intermittent
heating from a central active galactic nucleus (AGN; e.g. see
\citealt{Fabian03} for a recent review). However, it is important to
understand the origin of the large spread in $\dot{M}_{\rm X}$ within
the cluster population, as it also explains much of the scatter in the
luminosity--temperature relation \citep{Fabian94}.

\begin{figure}
\centering
\includegraphics[width=8cm]{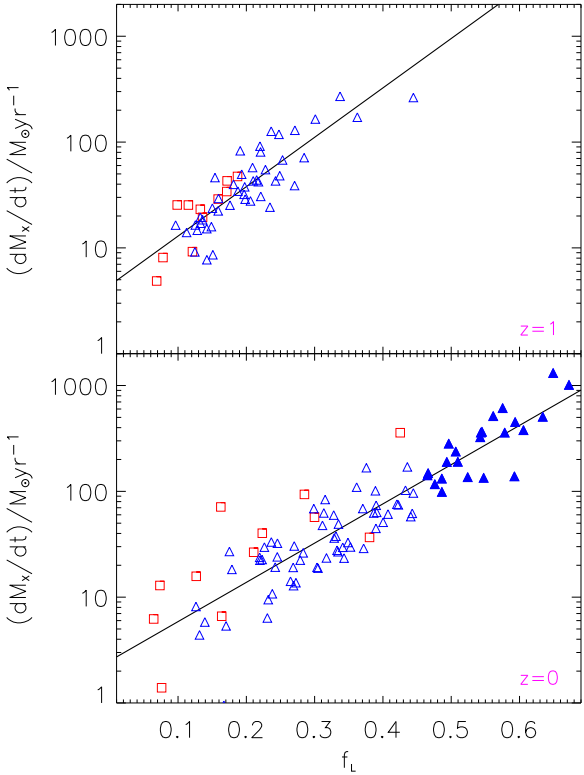}
\caption{Inferred mass deposition rate from X-ray emission,
$\dot{M}_{\rm X}$, versus X-ray concentration for regular (triangles)
and irregular (squares) clusters at $z=1$ and $z=0$. 
Filled triangles are regular clusters with $f_{\rm L}>0.45$.}
\label{fig:mdotxfl}
\end{figure}

We have measured $\dot{M}_{\rm X}$ for our clusters (within a fixed
physical radius of $r_{\rm core}=50 \hkpc$) and, as expected, found
that it is strongly correlated with $f_{\rm L}$, ranging from $\sim
1-900 M_{\odot}{\rm yr}^{-1}$ (Fig.~\ref{fig:mdotxfl}). Clusters with
the highest concentrations, $f_{\rm L}>0.45$, are regular and
typically have $\dot{M}_{\rm X}>100 M_{\odot}{\rm yr}^{-1}$. These
clusters could be called strong {\it cooling-flow} systems as they
most resemble the observational samples of the same name; again note
the absence of these objects at $z=1$.

While the median $f_{\rm L}$ decreases with redshift, the median
$\dot{M}_{\rm X}$ stays approximately constant. The lack of strong
{\it cooling-flow} clusters at high redshift is offset by the increase
in $\dot{M}_{\rm X}$ for individual systems, due to the ratio, $r_{\rm
core}/R_{500}$, being typically larger at higher redshift, thus
capturing more of the cluster's luminosity. Averaging over all
redshifts, we found that $\left< \dot{M}_{\rm X}\right> =35\pm 5
M_{\odot}{\rm yr}^{-1}$. 

Here, we do not attempt to address the issue of how much gas is {\it
actually} cooling down within our cluster cores. As discussed in
K2004b, our simulations currently lack the number of particles to
accurately follow the inward flow of the gas all the way down to low
temperature.  However, as we will demonstrate below, we find that the
large range in X-ray concentration/cooling-flow strength exhibited by
our clusters at low redshift is strongly dependent on the cluster's
larger-scale environment, i.e. whether it experienced a late-time
major merger or not.  So while the dynamics of a cooling core within a
given cluster may not be accurate, and requires further investigation,
our main (statistical) conclusions should hold as the simulation has
accurately followed the merger histories of the cluster population.

\subsection{Cooling flows, cool cores and dynamical state}

Besides their high core luminosity, cooling-flow clusters have
traditionally assumed to be dynamically-relaxed systems hosting a cool
core. Conversely, {\it non-cooling-flow} clusters with low core
luminosities are thought to host isothermal/warm cores and be
dynamically disturbed. This view-point was recently challenged by
\citet{McCarthy04} as being overly-simplistic, as observations of both
{\it cooling-flow} clusters with disturbed morphologies (e.g. Perseus)
and {\it non-cooling-flow} clusters (e.g. 3C 129) with relaxed
morphologies exist.  Our simulation lends some support to their
argument, as Fig.~\ref{fig:mdotxfl} shows. At $z=0$, irregular
clusters are found to have a large range in $f_{\rm L}$ (or
$\dot{M}_{\rm X}$), with one irregular cluster ($S_{\rm X}=0.14$)
having $\dot{M}_{\rm X}=358 M_{\odot}{\rm yr}^{-1}$. Conversely,
regular clusters can also have very low X-ray concentrations
($\dot{M}_{\rm X}<10 M_{\odot}{\rm yr}^{-1}$).  However,
statistically, the average regular cluster has a higher X-ray
concentration than an irregular cluster. This is because the X-ray
concentration is related to the dynamical history of the cluster, as
we will show below.

We showed in subsection~\ref{subsec:structparam} that $f_{\rm L}$ is
anti-correlated with the strength of the cool core, $f_{\rm
T}$, as measured from the projected temperature profile; clusters with
the coolest cores have more concentrated X-ray emission. However,
nearly all of our clusters, regular and irregular, have cool cores
($f_{\rm T}<1$; Figs.~\ref{fig:ftimages} \& \ref{fig:ftfl}). This is
in agreement with previous simulation work where the gas was allowed
to cool radiatively (e.g. \citealt{Motl04,Rowley04,Poole06}), where it
was found that cool cores are very hard to disrupt by mergers.

\begin{figure}
\centering
\includegraphics[width=8cm]{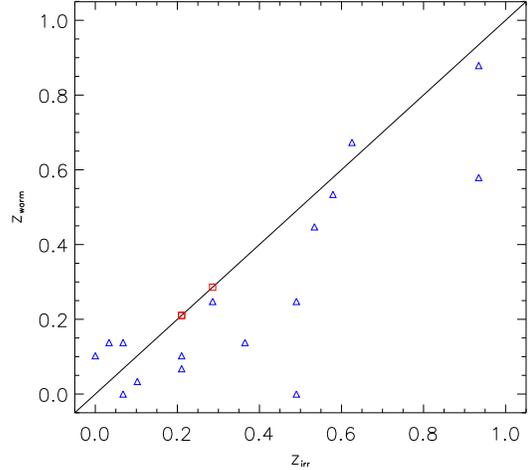}
\caption{Redshift when each cluster last had a warm core versus the
nearest redshift when it was irregular. Clusters that were regular
(irregular) at $z_{\rm warm}$ are shown as triangles (squares). The
solid line is $z_{\rm warm}=z_{\rm irr}$.}
\label{fig:lasthot}
\end{figure}

\begin{figure}
\centering
\includegraphics[width=8cm]{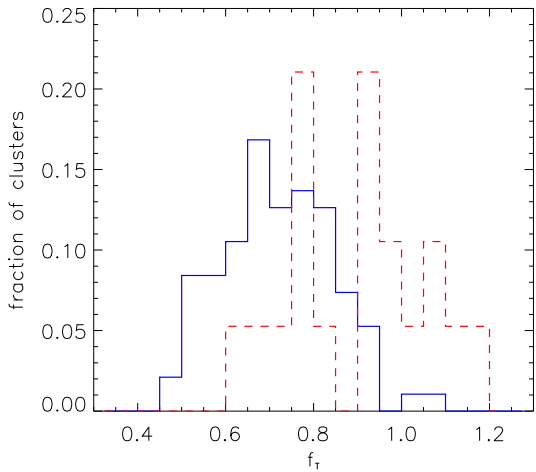}
\caption{Distribution of $f_{\rm T}$ values at $z=0$ (solid histogram), 
compared to the observational sample of \citet{Sanderson06}.}
\label{fig:ftdist}
\end{figure}

Warm (or non-cool; $f_{\rm T} \geq 1$) cores exist but are rare in our
simulation.  Given the number of outputs available, only one quarter
of the clusters were found to host a warm core since $z=1$, lasting at
most around 1 Gyr. Interestingly, clusters with warm cores nearly
always appear regular, even though the generation of a warm core
appears linked to the merger process. This is shown in
Fig.~\ref{fig:lasthot}, where we see a clear correlation between the
redshift when a cluster last had a warm core, $z_{\rm warm}$, against
the nearest redshift when it was irregular ($S_{\rm X}>0.1$), $z_{\rm
irr}$. It is unclear whether the cores are heated solely from the
gravitational interaction of the merger, or a contribution comes from
the feedback, which could also be triggered by a merger. Nevertheless,
the paucity of warm cores is at odds with the observational data at
low redshift. For example, \citet{Sanderson06} recently studied a
flux-limited sample of 20 clusters observed with {\it Chandra}, and
found only half of them to contain cool cores, even though the core
gas in the warm core clusters have cooling times significantly shorter
than a Hubble time. The discrepancy is illustrated clearly in
Fig.~\ref{fig:ftdist}, where we compare the distribution of $f_{\rm
T}$ values found in the CLEF simulation at $z=0$ with the
observational data of \citet{Sanderson06}. Although based on a limited
sample, the observations suggest there exists a bimodal distribution,
not present in the simulation. This suggests that our simulation is still 
missing a heating mechanism that could produce a larger fraction of 
warm cores, which again could be linked to AGN activity.

\subsection{Scatter in the L--T relation}

\begin{figure}
\centering
\includegraphics[width=8cm]{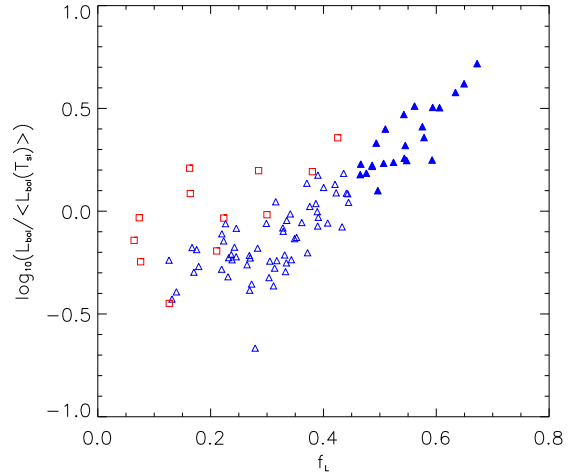}
\caption{Offset in luminosity from the mean $L-T$ relation versus
X-ray concentration for clusters at $z=0$. Triangles are regular
clusters and squares irregular clusters. Solid triangles are clusters
with the highest X-ray concentrations/X-ray-inferred mass deposition
rates.}
\label{fig:lbolfl}
\end{figure}

We now examine why there is a large scatter in the $L-T$ relation at
low redshift. Classically, it is thought that the scatter is related
to the dynamical histories of clusters. In particular, clusters with
the strongest cooling flows (which lie above the mean $L-T$ relation)
are believed to be in that state because they have not endured a major
merger in the recent past.  Our simulation supports this picture, as
will be demonstrated in the following two figures.  Firstly,
Fig.~\ref{fig:lbolfl} explicitly shows that the scatter in the $L-T$
relation is tightly correlated with the X-ray concentration (or mass
deposition rate) of a cluster. For regular systems, clusters with
higher X-ray concentrations lie above the mean relation, and those
with low X-ray concentrations below. Irregular clusters lie off this
correlation because $f_{\rm L}$ decreases due to the presence of a
second object (which also boosts the luminosity).

\begin{figure}
\centering
\includegraphics[width=8cm]{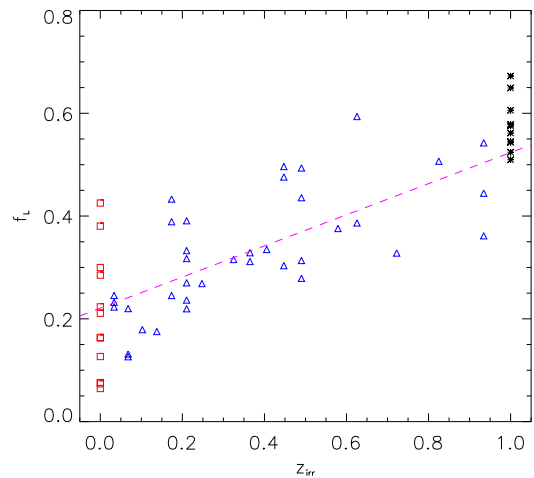}
\caption{X-ray concentration at $z=0$ versus the lowest redshift when
the cluster was irregular. Only clusters in our temperature-selected
sample at all available redshifts, $z<z_{\rm irr}$, are
plotted. Triangles are regular clusters today and squares irregular
clusters ($z_{\rm irr}=0$). Asterisks are clusters with $z_{\rm
irr}>1$, i.e. they did not experience a major merger between now
and $z=1$. The dashed line is a best-fit relation to the regular
clusters with $z_{\rm irr}<1$.}
\label{fig:lastirr}
\end{figure}

Secondly, in Fig.~\ref{fig:lastirr} we plot X-ray concentration at
$z=0$ versus the lowest redshift when each cluster experienced a major
merger.  Clusters which are not present in our temperature-selected
samples at all redshifts, $z<z_{\rm irr}$ are not plotted. Clearly
there is a strongly-positive correlation, demonstrating that the most
concentrated systems did not experience a major merger in the recent
past (the asterisks are those clusters with $z_{\rm irr}>1$).

An alternative mechanism for generating the scatter was proposed by
\citet{McCarthy04}, who used semi-analytic models of clusters with
preheating and cooling (but the effects of accretion and merging of
haloes were not included).  They suggested that the position of a
cluster on the luminosity--temperature relation was related to the
level of preheating it received: clusters that experienced higher
levels of preheating correspond to non-cooling-flow clusters (i.e. low
X-ray concentrations, here) and vice-versa. Similarly to
\citet{McCarthy04}, we tested whether the amount of feedback is
correlated to the strength of the cooling core.  Such an effect should
be seen through a trend of stellar mass fraction with $f_{\rm L}$, as
our feedback model injects energy approximately in proportion to the
star-formation rate. No trend is seen in our simulation, i.e. stellar
mass fractions are similar between clusters with low and high X-ray
concentrations.

It is clear, therefore, that the strong cooling-flow population exists
in our model at low redshift because of a lack of major merger
activity in such systems at $z<1$. The absence of strong cooling-flow
systems at higher redshift, responsible for the decrease in the $L-T$
scatter, can therefore be attributed to the increase in the merger
rate with redshift.

The absence of strong cooling-flow clusters at high redshift in our
model has important implications for cluster cosmology. Large samples
of X-ray clusters at high redshift are still in their infancy, 
although will start to become available over the next few years, such
as from the {\it XMM-Newton} Cluster Survey \citep{Romer01}.

If our prediction is correct, it will have both
positive and negative implications for cosmology. On the positive
side, the smaller scatter will allow for a simpler survey selection
function, with incompleteness effects being less of a problem. On the
negative side, there is a lack of very luminous objects, so the number
of high-redshift clusters above a given flux limit will be
considerably less, reducing the overall power for specific surveys to
constrain cosmological parameters. Interestingly, first observational 
results seem to support the lack of cooling-flow systems at high 
redshift \citep{Vikhlinin06}.

Another interesting point that our result throws up, is whether strong
cooling-flow clusters would exist in a universe with $\Omega_{\rm
m}=1$? In such a model, the merger rate would be expected to change
very little with redshift, so clusters today may not have had the time
to establish a strong cool core. In other words, the strongest
cooling-flow clusters only exist because of the freeze-out of
structure formation in a universe with sub-critical matter density.

\section{Conclusions}
\label{sec:conc}

In this paper, we presented the cluster population that forms within
the CLEF simulation, an $N$-body/hydrodynamics simulation of the
$\Lambda$CDM cosmology, with radiative cooling and energy feedback
from galaxies. Our cluster sample, with nearly one hundred $kT>2$ keV
objects at $z=0$ and sixty at $z=1$, is one of the largest drawn from
a single simulation. In this paper, we studied the demographics of the
cluster population out to $z=1$, focusing on the effects of dynamical
activity and the strength of cooling cores, and how the X-ray
properties of clusters depend on them. The Sunyaev--Zel'dovich
properties of the clusters may be found in a companion paper (da Silva
et al., in preparation). Our main conclusions are as follows:

\begin{itemize}

\item We quantified the amount of dynamical activity (major mergers)
within the cluster population, using a simple projected substructure
statistic, based on the observable X-ray surface-brightness
distribution. While there is no significant dependence of this
quantity, $S_{\rm X}$, with cluster temperature, it does increase with
redshift. The fraction of irregular, $S_{\rm X}>0.1$ clusters, shown
to be merging systems in the surface-brightness maps, increases from
around 10 per cent at $z=0$ to 20 per cent at $z=1$, thus constituting
a minority population at all redshifts.

\item The projected ICM temperature profile of regular clusters has a
generic shape at low and high redshift, decreasing in the centre (due
to radiative cooling) and beyond $0.2R_{500}$, due to the intrinsic
shape of the gravitational potential. Irregular clusters have flatter
profiles at large radii due to the presence of a second object which
compresses and heats the gas. The shape of the regular cluster profile
at $z=0$ is in good agreement with the recent study of cool core clusters
by \citet{Vikhlinin05a}. 

\item To quantify the core properties of our clusters, we defined two
simple (and observationally-measurable) structure parameters, $f_{\rm
T}$, which measures the core to maximum temperature ratio, and $f_{\rm
L}$ which measures the fraction of emission from within the core (the
X-ray concentration of the cluster). We found that the vast majority
of clusters contain cool cores  ($f_{\rm T}<1$) at all redshifts. This
is at odds with the observational data, at least at low redshift, where
only half of clusters contain cool cores \citep{Sanderson06}. The X-ray
concentration, $f_{\rm L}$, is anti-correlated with $f_{\rm T}$. The 
dispersion in $f_{\rm L}$ is large at $z=0$, but decreases with redshift
due to the absence of clusters with the highest values
(i.e. the strongest cooling cores).

\item The scaled entropy profile has an outer logarithmic slope of 0.9
and decreases all the way into the centre, with no evidence of a
flattened core.  The ratio of the normalisation at large radii, for
clusters at $z=1$ and $z=0$, is similar to that expected from the
gravitational-heating model ($S(T) \propto E^{-4/3}(z)$), but the
$z=1$ clusters have higher central entropy than at $z=0$.  Irregular
clusters have higher entropy profiles and regular clusters with strong
cooling cores have lower entropy profiles. The profile at $z=0$ (in
particular for the strong cooling core clusters) is in good agreement
with the recent observational data of \citet{Pratt06}.

\item Mass estimates of X-ray clusters, based on the hydrostatic
equilibrium equation, are around 20 per cent lower than the true
masses, even when spatial density and temperature information of the
ICM is known. As found by \citet{Rasia06}, the reasons for the 
discrepancy are X-ray temperature bias to low entropy gas and 
incomplete thermalisation of the gas.

\item The estimated mass versus spectroscopic-like temperature
relation at $z=0$ is only $\sim 10$ per cent higher than the
observed relation for $R_{500}$. Splitting the regular cluster sample
into those with weak and strong cooling cores makes little difference
to the properties of the relation, when the temperature is measured
outside the core.  Thus, details of the
mass--temperature relation should be insensitive to the cluster
selection procedure.

\item The mass--temperature relation evolves similarly to the
gravitational-heating model prediction, $M(T) \propto E^{-1}(z)$. The
scatter, $\Delta(\log M) \sim 0.08$, evolves very little with
redshift.

\item The luminosity--temperature relation has a large degree of
scatter at $z=0$, reflecting the large dispersion in X-ray
concentration of the clusters.  Excising the core emission reduces the
scatter considerably, although leads this to a relation that still has a 
higher normalisation than observed. Irregular clusters are not
systematically offset from the main relation. The
luminosity--temperature relation evolves negatively with redshift,
contrary to the gravitational-heating expectation, where $L(T) \propto
E(z)$. Excising the core reduces this negative evolution, with almost
self-similar evolution at very low redshift.

\item The scatter in the luminosity--temperature relation decreases
strongly with redshift, again due to the lack of strong cooling core
clusters at high redshift.  There is a positive correlation between
the X-ray concentration of the cluster and the redshift when it last had a
major merger, but apparently not between the X-ray concentration and 
the level of feedback experienced by the cluster. 
Thus, our results indicate that
the formation of a {\it cooling-flow} population of clusters at low
redshift is tied to the slow down in dynamical activity in the
$\Lambda$CDM model, allowing clusters in quieter environments to
develop a strong cooling core.

\end{itemize}

Our simulation is one of the first of a new generation that is able to
follow a substantial number of objects with reasonable resolution,
while attempting to include the vital physical processes that alter
the gravitationally-heated structure of the ICM: radiative cooling,
star formation and feedback. While our particular model can reproduce
many observed characteristic features of the cluster population,
particularly those with cool cores, we acknowledge that it has its
shortcomings. For example, it fails to completely quench the
overcooling of baryons into stars, it does not predict enough clusters
with warm cores, and it does not match the $L-T$ normalisation in
detail (being too high).

All these problems point to the need for an even more efficient
heating mechanism that reduces further the amount of cool gas in the
clusters, without destroying the already good agreement in cool core
clusters.  It may be possible that the problems could be overcome by
fine tuning the two feedback model parameters. However, it is
desirable to incorporate a more realistic physical model for feedback,
that is able to treat separately the effects from stars and black
holes (in our current model, the heating rate directly follows the
star-formation rate). The wealth of high-quality X-ray data that is
becoming available will undoubtedly help constrain the feedback
physics further, and thus allow more realistic cluster models to be
constructed.

\section*{Acknowledgements}

A.d.S. acknowledges support by CMBnet EU TMR network and Funda\c{c}\~{a}o
para a Ci\^{e}ncia e Tecnologia under contract SFHR/BPD/20583/2004,
Portugal. A.R.L.\ was supported by PPARC. A.R.L.\ thanks the Institute
for Astronomy, University of Hawai`i, for hospitality while this work
was completed.  The CLEF simulation was performed using 64 processors
on the Origin 3800 supercomputer at the CINES facility in Montpellier,
France; we wish to thank the support staff at CINES for their help. We
acknowledge partial support from CNES and Programme National de
Cosmologie. We would also like to thank Volker Springel for providing
a version of GADGET2 before its public release, Gabriel Pratt for
supplying observational data, and Etienne Pointecouteau and Arif Babul
for useful discussions.

\bsp

\label{lastpage}

\end{document}